%% file: paper_arxiv.tex
\documentclass[11pt]{article}

\usepackage{fullpage}
\usepackage[usenames,dvipsnames]{color}
\usepackage{amsmath}
\usepackage{amssymb}
\usepackage[numbers,square,sort&compress]{natbib}
\usepackage{listings}
\usepackage{graphicx}
\usepackage{float}

\usepackage{subfig}
\usepackage{tikz}
\usetikzlibrary{positioning,arrows,decorations,backgrounds,shapes}
\tikzstyle{expr}=[circle,draw=black]
\tikzstyle{dofexpr}=[rectangle,draw=black]
\tikzstyle{dep}=[thick,draw=black,-stealth']
\tikzstyle{dep2}=[thick,dashed,draw=red,-stealth']
\tikzstyle{dep3}=[thick,dotted,draw=blue,-stealth']

\newcommand{\cpp}{C++\ }

\newcommand{\R}{\mathbb{R}}
\newcommand{\Eqref}[1]{Eq.~\ref{#1}}

\newcommand{\Secref}[1]{Sect.~\ref{#1}}
\newcommand{\Secreff}[1]{Section~\ref{#1}}
\newcommand{\Figref}[1]{Fig.~\ref{#1}}

\input{symbol_defs}

\newcommand{\eg}{\mbox{\emph{e.g.\ }}}

\begin{document}

\title{Automating embedded analysis capabilities and managing software complexity in multiphysics simulation part I:  template-based generic programming}
\author{Roger P. Pawlowski, Eric T. Phipps\footnote{Corresponding author. Sandia National Laboratories, Optimization and Uncertainty Quantification Department, PO Box 5800 MS-1318, Albuquerque, New Mexico, 87185, USA. Tel: +1(505)284-9268, Fax: +1(505)845-7442, {\tt etphipp@sandia.gov}.}, and Andrew G. Salinger \\
Sandia National Laboratories\footnote{Sandia National Laboratories is a multi-program laboratory managed and operated by Sandia Corporation, a wholly owned subsidiary of Lockheed Martin Corporation, for the U.S. Department of Energy's National Nuclear Security Administration under contract DE-AC04-94AL85000.}}
\maketitle

\begin{abstract}
An approach for incorporating embedded simulation and analysis capabilities in complex simulation codes through template-based generic programming is presented.  This approach relies on templating and operator overloading within the \cpp language to transform a given calculation into one that can compute a variety of additional quantities that are necessary for many state-of-the-art simulation and analysis algorithms.  An approach for incorporating these ideas into complex simulation codes through general graph-based assembly is also presented.  These ideas have been implemented within a set of packages in the Trilinos framework and are demonstrated on a simple problem from chemical engineering.
\end{abstract}

\section{Introduction}
As computational algorithms, hardware, and programming languages have advanced over time, computational modeling and simulation is being leveraged to understand, analyze, predict, and design increasingly complex physical, biological, and engineered systems.  Because of this complexity, significant investments must be made, both in terms of manpower and programming environments, to develop simulation capabilities capable of accurately representing the system at hand.  At the same time, modern analysis approaches such as stability analysis, sensitivity analysis, optimization, and uncertainty quantification require increasingly sophisticated capabilities of those complex simulation tools.  Often simulation frameworks are not designed with these kinds of analysis requirements in mind, which limits the efficiency, robustness, and accuracy of the resulting analysis.  For example, sensitivity analysis, stability analysis, and optimization often need accurate estimates of derivatives of output quantities of interest with respect to simulation input data.  These derivatives can be estimated in a ``black-box'' fashion using some form of finite-differencing, however this is often both very inefficient and inaccurate.  This then limits the choice of algorithms that can be applied and reduces the number of quantities that can be analyzed.  It is more efficient and accurate to compute these derivatives analytically by embedding the calculation within the simulation code itself.  However because of the significant investments and complexity in these codes, it is often not practical to add the capabilities required by state-of-the-art analysis approaches after the fact.  Instead it is generally more attractive to design the programming environments with these kinds of capabilities in mind from the beginning.

Unfortunately this idea is very difficult to implement in practice.  The reality is the people who develop the simulation code capabilities are often different from those who develop the algorithmic analysis capabilities.  Each group often has very detailed, in-depth knowledge of their specific domain (e.g., complex physics simulation versus state-of-the-art uncertainty quantification algorithms), and it is usually impossible to be an expert in all of these domains.  Furthermore, the requirement to provide the extra information needed by these analysis algorithms (e.g., derivatives) often impedes the development of the simulation code, particularly early in the software development cycle (and thus there is often significant resistance to incorporate these capabilities early).  Finally, because of the long term investments in these codes, the analysis requirements can change over the life-cycle of the code and thus it may be impossible to incorporate from the beginning requirements that may arise later.

In this work, we describe an approach for building simulation code capabilities that natively support many types of analysis algorithms.  This approach leverages compile-time polymorphism and generic programming through \cpp templates to insulate the code developer from the need to worry about the requirements of advanced analysis, yet provides hooks within the simulation code so that these analysis techniques can be added later.  The ideas presented here build on operator overloading-based automatic differentiation techniques to transform a simulation code into one that is capable of providing analytic derivatives.  However we extend these ideas to compute quantities that aren't derivatives such as polynomial chaos expansions, floating point counts, and extended precision calculations.  In general any calculation that can be implemented in an operation-by-operation fashion can be supported, and the use of templates significantly simplifies the incorporation of many types of overloaded operators.  In addition to the template-based generic programming, we also present an approach for integrating these calculations into higher-level data structures appropriate for large-scale scientific computing based on a seed-compute-extract paradigm.  Finally we discuss a graph-based assembly approach for organizing the terms in the equations defining a complex multiphysics simulation and how to support generic programming and the seed-compute-extract paradigm.  Numerous challenges beyond those described here arise when applying these techniques to large-scale problems such as discretizations of multi-dimensional partial differential equations.  Our approach for applying template-based generic programming to these types of problems is the subject of a subsequent paper~\cite{Salinger:2011}.

While the concepts presented here are general and can be applied with a variety of software libraries, we base our discussion on a set of tools available within Trilinos~\cite{TrilinosTOMS} that implement these ideas.  In particular we describe an automatic differentiation package called Sacado~\cite{SacadoURL,Phipps2008LST} that provides the foundation for our template-based operator overloading approach.  Additionally we discuss a package called Stokhos~
\cite{StokhosURL} that generalizes the differentiation-based operator overloading in Sacado to a specific class of forward uncertainty quantification approaches reviewed in the next section.  We also present a library called Phalanx~\cite{PhalanxWebSite,DAG-TOMS2011} that implements the graph-based multiphysics assembly idea supporting template-based generic programming in complex physics simulation codes.  

Embedded analysis can be implemented using alternate techniques.  A successful approach used in the FEniCS \cite{FenicsOverview2007,LoggMardalEtAl2012a}, Life/FEEL++ \cite{Life2006a,Life2006b}, and Sundance \cite{Long2010} projects is to parse a symbolic representation of the problem written in a domain specific language (DSL) to generate the embedded quantities of interest.  The FEniCS project  requires the problem description to be specified in the unified form language \cite{UFL2012} and produces sensitivites through low level code generation.  The Life/FEEL++ project implements the symbolic front end using C++ expression templates and template metaprogramming techniques.  The Sundance project implements the symbolic front end using C++ expression objects (dynamic polymorhphism) and traverses the resulting expression tree to perform in-place differentiation.  While DSL approaches are quite successful, we feel that our approach provides some advantages in certain contexts.  When working in legacy code environments, the template-based approach allows for easy integration with the legacy components since the data interfaces are directly exposed to the applications.  In constrast, almost all DSL codes (except Sundance) constrain applications to implement the entire problem in the DSL, thus ruling out any complex operation that can't be described by the DSL (e.g., nonlinear elimination by a third party library \cite{Young03}).  A second advantage is that extensibility to new analysis types is easily supported.  Adding a new unexpected analysis capability such as polynomial chaos expansions to the back end parser in the symbolic approach can take considerable effort and expertise whereas with the template-based approach, the scalar type operations are overloaded regardless of the analysis capability.  The primary drawback to TBGP is that the flexibility provided by TBGP requires more effort to setup the initial machinery.  The DSL-based codes, on the other hand, typically have strong veritcal integration and have a much lower barrier for adoption.  A more detailed comparison of our assembly engine technology and DSLs can be found in \cite{DAG-TOMS2011}.

This paper is organized as follows.  We first provide a brief overview of several common simulation and analysis algorithms that are employed within modern scientific computing environments in \Secreff{sect:analysis}.  These algorithms motivate several derivative-based and non-derivative-based quantities that need to be computed within complex simulation codes.  We then review automatic differentiation techniques which provide the foundation for our work in \Secreff{sect:ad}, focusing in particular on operator overloading approaches in C++.  In \Secreff{sect:gen_oo} we generalize operator overloading for differentiation to compute more general quantities necessary in some forms of analysis which motivates the general template-based generic programming ideas presented in \Secreff{sect:generic}.  Then in \Secreff{sec:seed-compute-extract} we describe a seed-compute-extract paradigm that separates phases of the template-based generic programming approach into those that are specific to the kind of analysis chosen and those that are specific to the physical problem at hand.  In \Secreff{sec:complexity} we then discuss the generic graph-based assembly approach and how it incorporates template-based generic programming.  The approach is illustrated through a simple physical problem arising from chemical engineering where \cpp code for portions of the problem are provided.  Finally in \Secreff{sect:example} we provide a small example demonstrating these ideas for simulating and analyzing the chemical engineering problem from the previous section.  To perform these numerical calculations, several other Trilinos packages are leveraged including NOX~\cite{NoxURL} (nonlinear solver), LOCA~\cite{LocaURL,Salinger:2005p109} (stability and bifurcation analysis), Stokhos~\cite{StokhosURL} (uncertainty quantification), and Rythmos~\cite{RythmosURL} (time integration), all four of which are wrapped in a common interface through Piro~\cite{PiroURL} (analysis tools).  We then close with brief concluding remarks.

\section{Common Solution and Embedded Analysis Algorithms}\label{sect:analysis}
Before discussing our programming approach in detail, we first motivate the approach by presenting several common solution and analysis algorithms that often require significantly advanced simulation code capabilities.  As a model problem we will consider the general finite-dimensional differential-algebraic equation
\begin{equation} \label{eq:model}
  f(\dot{x},x,p) = 0, \quad\dot{x},x\in\R^n, \;\; p\in\R^m, \;\; f:\R^{2n+m}\rightarrow\R^n.
\end{equation}
Here $x$ is the unknown solution vector, $\dot{x}$ its time derivative, $p$ is a set of model parameters, and $f$ is the set of residual equations defining the model.

\subsection{Steady-state nonlinear solves}
Given a set of parameter values $p$, Newton-type methods are often employed to solve the steady-state version of \Eqref{eq:model} where $\dot{x}=0$, due to its quadratic convergence rate.  Given an initial guess $x_0$, the standard Newton's method involves solving the linear systems
\begin{equation}
  J(x_k,p) \Delta x_k = -f(x_k,p)
\end{equation}
for the update $\Delta x_k$ where for notational compactness we have dropped the dependence on $\dot{x} = 0$.  Here $x_{k+1} = x_k + \Delta x_k$, $k=0,1,\dots$ and $J = \partial f/\partial x$ is the Jacobian of $f$.  Thus the basic Newton's method requires evaluating the residual $f$ and Jacobian $J$ at an arbitrary point $(x,p)$.  There are many modifications of this algorithm to provide better global convergence properties (e.g., damping, line searches, and trust regions) as well as approaches that attempt to eliminate the need to compute the Jacobian directly (e.g., Broyden methods).  However we note the convergence and robustness of most of these methods are often significantly improved with accurate approximations of the Jacobian.

\subsection{Steady-state sensitivity analysis and optimization}
Once a steady-state solution $x^\ast$ has been computed via some nonlinear solver strategy, the sensitivity of the solution to the parameters $p$ can be computed directly through the implicit function theorem:
\begin{equation}
  \frac{\partial f}{\partial x}(x^\ast,p)\frac{dx^\ast}{dp} + \frac{\partial f}{\partial p}(x^\ast,p) = 0 \implies
  \frac{dx^\ast}{dp} = -J^{-1}(x^\ast,p)\frac{\partial f}{\partial p}(x^\ast,p).
\end{equation}
Often one is only interested in the sensitivity of a set of response functions $s^\ast = g(x^\ast,p)$, $g:R^{n+m}\rightarrow\R^q$, which can be computed by
\begin{equation}\label{eq:forward_sens}
  \frac{ds^\ast}{dp} = \frac{\partial g}{\partial x}(x^\ast,p)\frac{dx^\ast}{dp} + \frac{\partial g}{\partial p}(x^\ast,p) =
  -\frac{\partial g}{\partial x}(x^\ast,p)\left(J^{-1}(x^\ast,p)\frac{\partial f}{\partial p}(x^\ast,p)\right) + \frac{\partial g}{\partial p}(x^\ast,p).
\end{equation}
Thus the sensitivities $ds^\ast/dp$ can be computed via $m$ linear solves of the Jacobian $J$ with each right-hand-side given by $\partial f/\partial p_i$.  If all partial derivatives are computed analytically then the only error arising in the sensitivities comes from the error in solving these equations, which can often be controlled to some degree by the user through the choice of the linear solver algorithm.  This is to be contrasted with a finite differencing approach around the nonlinear solver, which entails $m$ nonlinear solves and introduces difficult to control finite difference truncation error.  Thus this approach is significantly more efficient, and often more robust.  It requires analytic evaluation of not only $f$ and $J$, but also $\partial f/\partial p$, $\partial g/\partial p$, and $(\partial g/\partial x)V$ for a given matrix $V$.

If the number of parameters $m$ is large but the number of response functions $q$ is small, then a better approach is to compute adjoint sensitivities, which are given by transposing \Eqref{eq:forward_sens}:
\begin{equation}\label{eq:adjoint_sens}
  \left(\frac{ds^\ast}{dp}\right)^T = -\left(\frac{\partial f}{\partial p}(x^\ast,p)\right)^T\left(J^{-T}(x^\ast,p)\left(\frac{\partial g}{\partial x}(x^\ast,p)\right)^T\right) + \left(\frac{\partial g}{\partial p}(x^\ast,p)\right)^T.
\end{equation}
This involves $q$ solves of $J^T$ where each right-hand-side is given by $\partial g_i/\partial x$.  It requires analytic evaluation of $f$, $J$, $\partial g/\partial x$, $\partial g/\partial p$, and $(\partial f/\partial p)^T W$ for an arbitrary matrix $W$.

With either the forward~\eqref{eq:forward_sens} or adjoint sensitivities~\eqref{eq:adjoint_sens}, a variety of gradient-based optimization algorithms can be applied for efficiently optimizing the quantity of interest $g$.

\subsection{Stability and bifurcation analysis} \label{sect:stability}
After a steady-state solution $x^\ast$ has been found, the transient stability of the solution can be investigated through linear stability analysis.  By considering infinitesimal perturbations away from the steady-state, simple analysis shows that the asymptotic stability can be understood  through the generalized eigenvalue problem
\begin{equation}\label{eq:eigen_sys}
  \lambda M z + J z = 0,
\end{equation}
where $M = \frac{\partial f}{\partial\dot{x}}(0,x^\ast,p)$ and $J = \frac{\partial f}{\partial x}(0,x^\ast,p)$.  If all of the eigenvalues $\lambda$ have a negative real part, then the solution is stable in a neighborhood of the solution.  If any eigenvalue has a positive real part, the solution is unstable and if any has zero real part, higher-order analysis is needed.  Thus in addition to the Jacobian $J$, derivatives with respect to the transient terms are needed.  Moreover, for an accurate determination of stability, it is critical to have exact analytic derivatives $J$ and $M$.

The location of the steady-state and its stability can be tracked as a function of the parameters $p$ through techniques such as pseudo-arclength continuation~\cite{keller79}.  Moreover parameter values at which a change in stability occurs, i.e. a bifurcation, can be solved for directly.  For example, a stable and unstable steady-state may collide as a parameter is increased yielding no steady-state for parameter values larger than where the bifurcation occurs, commonly called a saddle-node, fold, or turning-point bifurcation.  In this case one can show that \Eqref{eq:eigen_sys} has at least one solution with a zero eigenvalue $\lambda$, and thus $J$ has a non-trivial null-space ~\cite{Guckenheimer:83}.  Such a point can be computed directly by solving the following augmented system of equations~\cite{Govaerts:2000}:
\begin{equation} \label{eq:bif}
  \begin{split}
    f(x,p) = 0, \\
    \sigma(x,p) = 0,
  \end{split}
\end{equation}
where
\begin{equation} \label{eq:sigma_def}
\sigma = -u^TJv, \quad
\begin{bmatrix}
  J   & a \\
  b^T & 0
\end{bmatrix}
\begin{bmatrix}
  v \\
  s_1
\end{bmatrix} =
\begin{bmatrix}
  0 \\
  1
\end{bmatrix}, \quad
\begin{bmatrix}
  J^T & b \\
  a^T & 0
\end{bmatrix}
\begin{bmatrix}
  u \\
  s_1
\end{bmatrix} =
\begin{bmatrix}
  0 \\
  1
\end{bmatrix},
\end{equation}
and $a$ and $b$ are given vectors.  This system has a unique solution at a saddle-node bifurcation for almost all choices of $a$ and $b$, under suitable conditions~\cite{Govaerts:2000}.  Applying a Newton-type nonlinear solver to \Eqref{eq:bif} requires calculation of the derivatives $\partial\sigma/\partial x$ and $\partial\sigma/\partial p$.  From \Eqref{eq:sigma_def} it can be shown that
\begin{equation}\label{eq:sigma_derivs}
  \frac{\partial\sigma}{\partial x} = -u^T\frac{\partial}{\partial x}(Jv), \quad \frac{\partial\sigma}{\partial p} = -u^T\frac{\partial}{\partial p}(Jv).
\end{equation}
Thus this method requires calculation of various second derivatives.  As with stability analysis, it is critical to have an accurate approximation of the Jacobian $J$ since it appears directly in the equations defining the bifurcation.  It is less critical to accurately evaluate the second derivatives appearing in \Eqref{eq:sigma_derivs} since they are only used in the Newton solver method, but as discussed above, robustness and convergence are often improved if these derivatives are computed accurately.

\subsection{Transient analysis}
\label{sec:transient_analysis}
There are a variety of explicit and implicit time integration strategies for computing numerical solutions to the transient problem~\eqref{eq:model} with suitable initial conditions.  For stiff or differential-algebraic problems, implicit backward differentiation formulas (BDF)~\cite{HairerWannerI,HairerWannerII} are often applied.  These methods approximate $\dot{x}$ through the multistep formula
\begin{equation}\label{eq:bdf}
  h_n \dot{x}_n = \sum_{i=0}^q \alpha_i x_{n-i}
\end{equation}
where $q$ is the order of the method, $h_n$ is the time step size at time step $n$, and $x_j$, $\dot{x}_j$ represent the solution and its time derivative respectively at time step $j$.  Substituting \Eqref{eq:bdf} in \Eqref{eq:model} we obtain the following nonlinear system to solve for $x_n$:
\begin{equation}
  f\left(\frac{1}{h_n}\sum_{i=0}^q \alpha_i x_{n-i}, x_n, p\right) = 0.
\end{equation}
Solving this system for $x_n$ with a Newton-type method requires solving linear systems of the form
\begin{equation}
 \left(\frac{\alpha_0}{h_n}\frac{\partial f}{\partial\dot{x}} + \frac{\partial f}{\partial x}\right) \Delta\x_n = - f.
\end{equation}
Thus derivatives similar to the eigensystem \Eqref{eq:eigen_sys} must be computed.

\subsection{Uncertainty quantification}\label{sec:uq}
The last form of analysis we will consider is one where the parameters $p$ appearing in \Eqref{eq:model} have uncertainty associated with them.  Typically this uncertainty manifests through lack of knowledge as to what the values of the parameters in the system model~\eqref{eq:model} should be (so called epistemic uncertainty), or represents some variability in realizations of those parameters (so called aleatory uncertainty).  For analysis purposes, the uncertainty must be represented in some mathematical fashion, such as bounds or intervals in the first case, or random variables with prescribed probability distributions in the second.  Then the uncertainty quantification problem is to estimate a representation of the uncertainty of the corresponding solution $x$ to \Eqref{eq:model} (either steady or transient).  There are a variety of computational approaches for this in the literature, but recently response surface methods have become popular.  These methods compute a response surface approximating the mapping $x(p)$, which can then be used to approximate the uncertainty in $x$.  For example, $x(p)$ can be sampled to compute bounds on $x$ for epistemic analysis or probabilities and statistics in an aleatory analysis.

One response surface approximation method in particular that has seen significant attention in the literature is one based on approximating $x(p)$ in terms of orthogonal polynomials, often called polynomial chaos~\cite{Wiener:1938p989,Ghanem:1990p7167,Ghanem_Spanos_91,Xiu:2002p919}.  For simplicity, consider the steady-state version of \Eqref{eq:model} where to represent the uncertainty in the parameters $p$ we replace them with random variables $\xi$ with a known probability measure $\mu$ and density function $\rho$.  More precisely, assume $(\Omega,\mathcal{B},P)$ is a given probability space, with sample space $\Omega$, $\sigma$-algebra $\mathcal{B}$ representing admissible events, and probability measure $P:\mathcal{B}\rightarrow[0,1]$, then $\xi:\Omega\rightarrow\R^m$, $\mu = P\circ\xi^{-1}$ is the measure of $\xi$, and $\rho = d\mu/d\lambda$ where $\lambda$ is Lebesgue measure on $\R^m$.  Then formally $x(\xi)$ can be written as the following sum:
\begin{equation}\label{eq:pc_sum}
  x(\xi) = \sum_{i=0}^\infty x_i \psi_i(\xi)
\end{equation}
where $\{\psi_i: i=0,1,2,\dots\}$ is a family of polynomials orthogonal with respect to the measure $P$:
\begin{equation}\label{eq:orthog}
  \int_\Omega \psi_i(\xi(\omega))\psi_j(\xi(\omega))dP = \int_{\R^m} \psi_i(\xi)\psi_j(\xi)d\mu = \int_{\R^m}\psi_i(y)\psi_j(y)\rho(y)dy = \langle\psi_i^2\rangle\delta_{ij}.
\end{equation}
and where $\langle\cdot\rangle = \int\cdot\rho dy$.  The convergence of this sum is in the $L^2_P(\Omega)$ sense.  The coefficients $x_i$ are generalized Fourier coefficients, and thus \Eqref{eq:pc_sum} is often called a spectral representation of $x(\xi)$.

For computational purposes, the sum~\eqref{eq:pc_sum} must be truncated at some finite order $N$ and the coefficients $x_i$ must be approximated.  Owing to the orthogonality relation~\eqref{eq:orthog}, we have
\begin{equation}
  x_i = \frac{1}{\langle\psi_i^2\rangle}\int_{\R^m} x(y)\psi_i(y)\rho(y)dy
\end{equation}
which can be approximated by a multi-dimensional quadrature rule $\{(w^k,y^k): k=0,\dots,Q\}$ defined by $\rho$:
\begin{equation}
  x_i \approx \frac{1}{\langle\psi_i^2\rangle}\sum_{k=0}^Q w^k x(y^k)\psi_i(y^k).
\end{equation}
For each $k$, $x^k = x(y^k)$ is computed by solving $f(x^k,y^k) = 0$.  This is the so-called non-intrusive spectral projection (NISP)~\cite{Reagan:2003p920} method.  In high dimensions $m$, an efficient quadrature rule is difficult to find for a general random vector $\xi$.  When the components are independent, sparse-grid quadrature rules~\cite{Novak:1996p4613} are the most effective.  Thus a sequence of $Q+1$ steady-state solves are necessary for each quadrature point realization $y_k$ of the random parameters $\xi$.

Another approach for approximating the coefficients $x_i$, the so-called stochastic Galerkin approach~\cite{Ghanem:1990p7167,Ghanem_Spanos_91}, is to solve the equations
\begin{equation}\label{eq:sg_resid}
  F_i \equiv \frac{1}{\langle\psi_i^2\rangle}\int_{\R^m} f(\hat{x}(y),y)\psi_i(y)\rho(y)dy = 0, \quad i=0,\dots,N,
\end{equation}
where $\hat{x}(y) = \sum_{i=0}^N x_i\psi_i(y)$.  This generates a new, fully-coupled set of nonlinear equations for all of the $x_i$:
\begin{equation}
  F(X) = 0, \quad X = [x_0,\dots,x_N]^T, \quad F = [F_0,\dots,F_N]^T.
\end{equation}
To solve these equations with Newton's method, calculation of the Jacobian matrix $\partial F/\partial X$ is necessary.  From \Eqref{eq:sg_resid} one can show
\begin{equation}
  \frac{\partial F_i}{\partial x_j} = \frac{1}{\langle\psi_i^2\rangle}\int_{\R^m}\frac{\partial f}{\partial x}(\hat{x}(y),y)\psi_i(y)\psi_j(y)\rho(y)dy \approx \sum_{k=0}^N J_k \frac{\langle \psi_i\psi_j\psi_k\rangle}{\langle\psi_i^2\rangle}
\end{equation}
where
\begin{equation}\label{eq:sg_jac}
  J_k = \frac{1}{\langle\psi_k^2\rangle}\int_{\R^m}\frac{\partial f}{\partial x}(\hat{x}(y),y)\psi_k(y)\rho(y)dy
\end{equation}
are the polynomial chaos coefficients of the Jacobian operator $\partial f/\partial x$.  Unlike the NISP method above, this method is intrusive as it requires the formulation and solution of this new nonlinear system $F(X)=0$, and thus requires significant software infrastructure to implement.  In particular the simulation code must implement a method for computing the stochastic Galerkin residuals $F_i$ and Jacobians $J_k$.  The primary reason for doing so is that typically the number $N$ of unknown coefficients $x_i$ is significantly smaller than the number $Q$ of quadrature points $y_k$, which can lead to computational savings in some cases~\cite{Elman:2011p6562}.

\section{Automatic Differentiation Through Operator Overloading} \label{sect:ad}
As we have seen above, advanced simulation and analysis algorithms require accurate and efficient evaluation of a variety of first and higher derivatives, as well as evaluation of other quantities such as polynomial chaos expansions.  For the reasons discussed in the introduction, it is not practical to expect a simulation code implementing the evaluation of $f$ in \Eqref{eq:model} to provide implementations of all of these calculations.  Thus techniques for automatically generating this information are necessary.  The conceptual foundation for our approach to automating these calculations lies with automatic differentiation techniques, specifically through operator overloading.

\subsection{Introduction to automatic differentiation}\label{sect:ad_intro}
Automatic differentiation (AD) is a well-known set of techniques for transforming a given computation implemented in a programming language, into one that computes derivatives of that computation.  For an in-depth introduction to AD techniques we refer the reader to~\cite{Griewank2000EDP} and the references contained within.  Briefly, automatic differentiation works by recognizing that any computation implemented in a programming language must be a composition of simple arithmetic operations (addition, subtraction, multiplication, and division) and standard mathematical functions (sine, exponential, etc\ldots).  All of these have known formulas for computing their derivatives, and thus computer code for evaluating the derivative of the computation can be generated by combining these formulas with the chain rule of basic differential calculus.  Automatic differentiation is just a method for automating this process, and the end result is computer code that computes the derivative of the calculation at a given set of inputs.  Most realistic computations also involve branches, e.g., conditionals and loops, which can be handled by a variety of mechanisms.  The simplest is to only compute the derivative of the branch that was evaluated at the given set of inputs upon which the original calculation was based (this means the choice of branches within the derivative code is only based on the values of the original calculation and not its derivatives).  For more detail on approaches for dealing with branches, discontinuities, and iteration, see~\cite{Griewank2000EDP}.

There are different modes of automatic differentiation that accumulate derivatives in different orders, namely the forward and reverse modes, as well as modes to compute higher order derivatives (e.g., multivariate tensors and univariate Taylor series).  For example, if the calculation to be differentiated is represented mathematically as $y = f(x)$ where $x\in\R^n$ and $y\in\R^m$, then given a point $x_0\in\R^n$ and an arbitrary matrix $V\in\R^{n\times p}$, the forward mode of AD computes the derivative (see the discussion below on seed matrices as to why this form is chosen)
\begin{equation}\label{eq:forward_ad}
  \frac{dy}{dx}V = \frac{df}{dx}(x_0)V = \left.\frac{d}{dz}f(x_0 + Vz)\right|_{z=0}.
\end{equation}
Forward AD implements this calculation by computing derivatives with respect to $z$ at each stage of the calculation using rules such as those from Table~\ref{tab:forward_ad_rules}, propagated forward starting from the initialization $dx/dz = V$.  Owing to the directional derivative nature of formula~\eqref{eq:forward_ad}, this method is often called tangent propagation.  One can then easily show~\cite{Griewank2000EDP} that the cost of the resulting derivative calculation is bounded by
\begin{equation}
  \mbox{cost}\left(\frac{dy}{dx} V\right) \sim (1 + 1.5 p)\mbox{cost}(f)
\end{equation}
where $\mbox{cost}(f)$ is the cost to evaluate the function $f$.  Since all of the columns of $V$ are usually propagated simultaneously, the cost of forward mode AD seen in practice is often significantly less~\cite{Bartlett2006ADo}.
\begin{table}
  \begin{center}
    \caption{Forward AD differentiation rules for several scalar intermediate operations.  Overdots denote differentiation with respect to a single independent variable $z$.}
    \label{tab:forward_ad_rules}
    \bigskip
    \begin{tabular}{|l|l|}
      \hline
      \multicolumn{1}{|c|}{Operation} & \multicolumn{1}{|c|}{Forward AD rule} \\
      \hline
      $c = a\pm b$ & $\dot{c} = \dot{a}\pm \dot{b}$ \\
      $c = a b$ & $\dot{c} = a\dot{b} + \dot{a} b$ \\
      $c = a/b$ & $\dot{c} = (\dot{a} - c\dot{b})/b$ \\
      $c = a^r$ & $\dot{c} = r a^{r-1}\dot{a}$ \\
      $c = \sin(a)$ & $\dot{c} = \cos(a)\dot{a}$ \\
      $c = \cos(a)$ & $\dot{c} = -\sin(a)\dot{a}$ \\
      $c = \exp(a)$ & $\dot{c} = c\dot{a}$ \\
      $c = \log(a)$ & $\dot{c} = \dot{a}/a$ \\
      \hline
    \end{tabular}
  \end{center}
\end{table}

Similarly, given an arbitrary matrix $W\in\R^{m\times q}$, reverse AD computes
\begin{equation}
  W^T \frac{dy}{dx} = W^T \frac{df}{dx}(x_0) = \left.\frac{d}{dx}(W^T f(x))\right|_{x=x_0}
\end{equation}
by computing derivatives of $z = W^T y$ with respect to each intermediate quantity using rules such as those from Table~\ref{tab:reverse_ad_rules}, propagated backward starting from the initialization $dz/dy = W^T$.  This mode is often referred to as adjoint or gradient propagation.  One can show the cost is given by
\begin{equation}~\label{eq:reverse_ad_cost}
  \mbox{cost}\left(W^T \frac{dy}{dx}\right) \sim 4 q\;\mbox{cost}(f).
\end{equation}
Thus for a scalar valued calculation ($m=1$), its gradient ($q=1$) can be computed in the cost of about four function evaluations, regardless of the number of independent variables.  Note that since reverse AD requires propagating derivatives backwards through the calculation, some facility for reversing the code evaluation is necessary.  Typically this is accomplished by building a data structure storing the computational graph of the calculation generated through a forward pass.  This data structure stores the values of all of the intermediate operations, some mechanism for computing the derivative of each operation (either encoding some functional representation of the operation or a partial derivative computed in the forward pass), and connectivity representing the arguments of each operation.  This data structure can then be traversed in reverse order to accumulate the derivatives.  However because of the overhead associated with traversing this data structure, often the cost of reverse mode AD is somewhat higher than the bound~\eqref{eq:reverse_ad_cost} above~\cite{Bartlett2006ADo}.
\begin{table}
  \begin{center}
    \caption{Reverse AD rules for several scalar intermediate operations.  Overbars over each intermediate variable denote differentiation of a single dependent variable $z$ with respect to that intermediate variable, e.g., $\bar{a} = dz/da$.  For examples putting these rules together to differentiate a function, we refer the reader to~\cite{Griewank2000EDP}.}
    \label{tab:reverse_ad_rules}
    \bigskip
    \begin{tabular}{|l|ll|}
      \hline
      \multicolumn{1}{|c|}{Operation} & \multicolumn{2}{|c|}{Reverse AD rule} \\
      \hline
      $c = a+b$ & $\bar{a}=\bar{c}$ & $\bar{b}=\bar{c}$ \\
      $c = a-b$ & $\bar{a}=\bar{c}$ & $\bar{b}= -\bar{c}$ \\
      $c = a b$ & $\bar{a}=\bar{c}b$ & $\bar{b}=\bar{c}a$ \\
      $c = a/b$ & $\bar{a}=\bar{c}/b$ & $\bar{b}= -\bar{c}c/b$ \\
      $c = a^r$ & $\bar{a}=\bar{c}r a^{r-1}$ & \\
      $c = \sin(a)$ & $\bar{a}=\bar{c}\cos(a)$ & \\
      $c = \cos(a)$ & $\bar{a}= -\bar{c}\sin(a)$ & \\
      $c = \exp(a)$ & $\bar{a}=\bar{c}c$ & \\
      $c = \log(a)$ & $\bar{a}=\bar{c}/a$ & \\
      \hline
    \end{tabular}
  \end{center}
\end{table}

The matrices $V$ and $W$ are referred to as seed matrices, and the user is allow to choose them to suit their purposes.  For example if the whole Jacobian matrix $dy/dx$ is desired, forward or reverse mode can be used with the corresponding seed matrix set to the identity.  Similarly, the seed matrix can be set to a single vector to compute Jacobian-vector or Jacobian-transpose-vector products with significantly less cost than computing the whole matrix.  Finally, the forward and reverse modes can be applied recursively to compute a variety of higher derivatives.

\subsection{Automatic differentiation software tools}

There are a number of software tools available that implement the various modes of automatic differentiation discussed above in a variety of programming languages.  Most tools can be put into two categories:  those based on source transformation and those based on operator overloading.  Source transformation tools operate by reading the source code to be differentiated, parsing it, building an internal representation of the calculation, applying the AD rules discussed above, generating derivative code and writing this code out to disk.  The derivative code can then be compiled and linked into an application using standard compiler tools.  It is up to the user to write additional code that calls the differentiated routines and incorporates the derivatives in whatever way necessary.  This approach has been most successful for languages such as Fortran through tools such as ADIFOR~\cite{Bischof1996A2A}, Tapenade~\cite{Hascoet2004TUG}, and OpenAD~\cite{Utke2008OAM}, but has also been used for C (e.g., ADIC~\cite{Bischof1997AAE}) and recently some \cpp (e.g., OpenAD and TAC++~\cite{Vossbeck2008DaF}).  This approach is ideal in the sense that it provides the automatic differentiation tool a global view of the code to be differentiated, allowing optimization of the resulting derivative calculation.

However the source transformation approach has seen less use with the \cpp language, primarily due to the complexity of parsing and supporting all of the features of ANSI C++.  Instead the simpler approach of leveraging the native classing, operator overloading, and templating language features of \cpp is more popular.  This approach is library based, whereby new data types are created to store both function values and derivative values, and overloaded versions of the arithmetic and math functions are provided  that implement the derivatives of those operations operating on these new data types using rules such as those from Tables~\ref{tab:forward_ad_rules} and~\ref{tab:reverse_ad_rules}.  Then the floating point data type in a given calculation (e.g., {\tt float} or {\tt double}) is replaced by the corresponding AD data type.  The compiler then applies its operator overloading resolution rules to replace the original function calls with the differentiated versions.  Templating can be used (although is not required) to automate the necessary type replacement.  As with source transformation, the user must provide additional code to call the differentiated calculation, as well as code to initialize the AD data types and extract the resulting derivatives.

There are numerous tools available implementing this approach, including ADOL-C~\cite{Griewank1996AAC}, FAD~\cite{Aubert2001ADi,Aubert2002ETa}, FADBAD~\cite{Bendtsen1996FaF} and Sacado~\cite{Phipps2008LST}.  All of the automatic differentiation modes can be implemented this way, and it can be applied to any language that supports derived types and operator overloading (including Python and MATLAB).  However the disadvantage of this approach is that the resulting derivative calculation can be less efficient than the one resulting from source transformation.  This is partly due to the fact that operator overloading-based tools can only manipulate individual operations and thus don't have an opportunity to optimize the whole derivative calculation, but is mainly due to the extra overhead operator overloading introduces.  In essence, each operation in the original calculation is replaced by a function call, which adds overhead.  Moreover, \cpp requires that the overloaded operators return copies of the object created within them representing the result of the differentiated operation, which is then copied into the object to store the result in the calling code.  This again adds overhead.  Some of this overhead can be automatically eliminated by the compiler through aggressive function inlining and optimization.  However more recent AD tools such as FAD and Sacado employ expression templates to trick the compiler into eliminating virtually all of this overhead.

\subsection{Expression templates}

Expression templates are a general set of techniques that make operator overloading more efficient~\cite{VelhuizenET}.  Expression templates have been used in numerous contexts such as the Blitz++ vector library~\cite{Veldhuizen:1998}, the LLANO~\cite{Kirby:2003gz} and Blaze~\cite{Iglberger:2012hb} dense linear algebra libraries, and the Playa sparse linear algebra package~\cite{Playa2012}, however for concreteness we will describe them in the context of AD.  Essentially they work by creating a new data type for each expression in the calculation.  This data type is a tree that encodes the structure of the expression, the types of operations in the expression, and their arguments.  Instead of differentiating each individual operation directly, each overloaded operator appearing in an expression adds a node to the tree to represent the operation by combining the expression trees for its arguments.  Thus the arguments and return type of the operator are expressions instead of AD data types.  An overload of the assignment operator must then be provided for the AD data type with an arbitrary expression as the right-hand-side which then recursively differentiates the entire expression.  The type of each operation in the expression can be specified by a template parameter, and each node in the tree must only store a reference to the expressions representing its arguments.  Thus the creation of the expression template involves no runtime calculation, and the entire expression object can usually be eliminated through aggressive compiler optimizations.  This eliminates much of the overhead of operator overloading by eliminating the creation and copying of temporary objects and fusing the differentiation of all of the operations within an expression.  This latter optimization allows a single loop to be created iterating over all of the components of the derivative array, instead of having separate loops for for each operation.  Moreover if the length of this array (i.e., the number of independent variables) is known at runtime, this loop can be unrolled by the compiler.

\begin{figure}[H]
  \centering
  \includegraphics[scale=0.5]{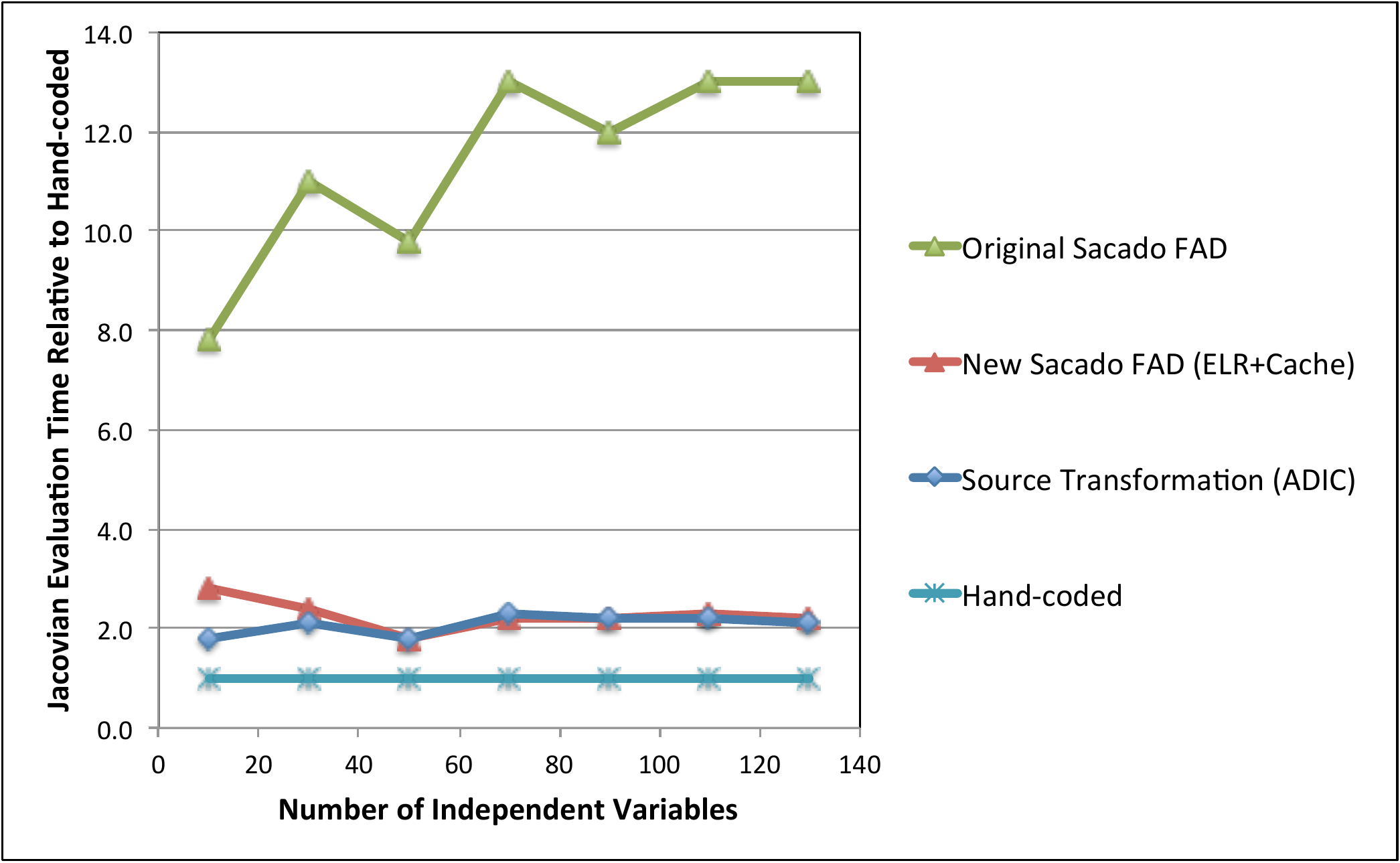}
  \caption{Comparison of Jacobian evaluation times for a simple vector PDE using several approaches:  the original Sacado FAD AD tools in Sacado based on the original FAD library, the new Sacado FAD tools incorporating expression-level reverse mode and caching, source transformation with ADIC~\cite{Bischof1997AAE}, and a hand-coded, optimized Jacobian evaluation.  Times are relative to the hand-coded approach and all cases were implemented using recent GNU C/C++ compilers (4.6.3) with standard -O3 optimization flags.  In all cases the Jacobian values agree to machine precision.  The comparison to source transformation demonstrates that for a sufficiently large number of independent variables, the overhead traditionally associated with operator overloading can be eliminated by the new expression template implementations.}
\label{fig:fad}
\end{figure}

FAD~\cite{Aubert2001ADi,Aubert2002ETa} was the first AD library we are aware of that implemented expression templates, and these ideas were later incorporated into the Sacado~\cite{Phipps2008LST} AD library in Trilinos.  Sacado provides further refinements of the expression template ideas in FAD to make them more efficient~\cite{Phipps2012LST}.  In particular, Sacado implements an expression level reverse mode where each operation is differentiated by the reverse mode AD approach.  This reduces the number of floating point operations by recognizing that an expression often has many inputs, but only one output.  The expression-level reverse mode computes the gradient of this output with respect to the expression inputs, and then applies the chain rule to obtain the derivatives with respect to the independent variables.  Furthermore, Sacado can cache the value of each operation in the expression tree for use in the subsequent derivative evaluation.  This trades a small amount of data storage required of the expression for a reduction in the number of floating point operations.  This is particularly important when nesting AD types for higher derivatives or for other kinds of non-derivative analysis (such as polynomial chaos expansions) discussed below.  These optimizations together with modern optimizing compilers significantly improve the performance of the forward AD tools~\cite{Phipps2012LST}, virtually eliminating the overhead typically associated with operator overloading (see Figure~\ref{fig:fad}).  In addition to forward mode AD, Sacado also provides reverse-mode AD tools called RAD~\cite{Gay2005SDf,Bartlett2006ADo}.  Both the forward and reverse mode AD data types can be combined to compute a variety of higher derivatives through a general template mechanism discussed below.

\section{Generalized Operator Overloading For Embedded Analysis}\label{sect:gen_oo}

The operator overloading approaches discussed above provide a relatively simple mechanism for generating analytic derivatives of arbitrarily complex calculations.  This idea can be extended in a rather straight forward manner to compute other kinds of quantities that aren't derivatives.  For example, several packages are available that compute floating point operations in higher than double precision (e.g.,~\cite{GmpUrl}).  In \cpp these packages work by implementing a new data type that stores the floating point value for each operation, and overloads all of the same operations discussed above to compute these operations in extended precision.  Similarly Sacado (and likely other AD packages) provides a simple floating point operation counting class that, along with its overloaded operators, can count the number of floating point operations (as well as other operations such as copies and assignments) in a given calculation.

Finally, operator overloading can be used to generate information needed for the uncertainty quantification methods discussed in \Secref{sec:uq}.  For example, the stochastic Galerkin method requires the evaluation of the stochatic Galerkin residual equations~\eqref{eq:sg_resid}.  This calculation can be implemented by creating a data type storing the polynomial chaos coefficients for a given intermediate quantity used in the evaluation of the residual function $f$.  Then overloaded operators can be written that compute these coefficients for each type of intermediate operation.  For example, assume $a$ and $b$ are two intermediate quantities in the calculation for $f$ and assume by way of induction that their polynomial chaos expansions
\begin{equation}
  a(\xi) = \sum_{i=0}^N a_i\psi_(\xi), \quad b(\xi) = \sum_{i=0}^N b_i\psi_i(\xi)
\end{equation}
have already been computed.  We wish to compute the chaos expansion
\begin{equation}
  c(\xi) = \sum_{i=0}^N c_i\psi_(\xi),
\end{equation}
where $c = \varphi(a,b)$ is given by a standard elementary operation of $a$ and $b$.  If $c = a\pm b$ then clearly $c_i = a_i\pm b_i$, $i=0,\dots,N$.  If $c=ab$, then by orthogonality of the basis polynomials we have
\begin{equation}
  c_k = \sum_{i,j=0}^N a_ib_j\frac{\langle\psi_i\psi_j\psi_k\rangle}{\langle\psi_k^2\rangle}, \quad k=0,\dots,N.
\end{equation}
Similarly if $c = a/b$, then we obtain the following linear system of equations to solve for the coefficients $c_i$:
\begin{equation}
  \sum_{i,j=0}^N c_i b_j\frac{\langle\psi_i\psi_j\psi_k\rangle}{\langle\psi_k^2\rangle} = a_k, \quad k=0,\dots,N.
\end{equation}
Computing stochastic Galerkin projections of transcendental operations is more challenging, and several approaches have been investigated~\cite{Debusschere:2004p96}.  A relatively simple and robust approach is quadrature inspired by non-intrusive spectral projection:
\begin{equation}
  c_k \approx \frac{1}{\langle\psi_k^2\rangle} \sum_{j=0}^Q w^k \varphi(a(y^k),b(y^k))\psi_k(y^k), \quad \quad k=0,\dots,N.
\end{equation}
Sacado, together with the stochastic Galerkin package Stokhos~\cite{StokhosURL} implement these approaches to provide an operator overloading mechanism for generating stochastic Galerkin residual values.

\section{Template-based Generic Programming and Compile-time Polymorphism} \label{sect:generic}

As we have seen, a myriad of analysis capabilities can be incorporated into complex simulation codes through the creation of specialized types, which we will call scalar types, and corresponding overloaded operators.  Applying each of these to the code requires replacing the basic floating point type with a new scalar type.  Since this conversion must happen for many different types, we advocate a templating approach where the scalar type is replaced by a template parameter.  When this template code is instantiated on the original floating point type, the original code is obtained (and thus no performance is lost in the original calculation), and when it is instantiated on each new scalar type, the corresponding overloaded evaluation is obtained.  This simplifies extending the AD library to support new modes since only a new scalar data type and corresponding overloaded operators must be provided and the template code can be easily instantiated on this new type.  Furthermore the code developer only needs to develop, test, and maintain one templated code base.  Finally the implementation of the Sacado scalar data types and overloaded operators are themselves templated on the scalar type to allow nesting of types.  For example, by replacing the floating point type in a forward mode AD evaluation with the forward mode AD data type, a second derivative of the form
\begin{equation}
  \frac{d}{dx}\left(\frac{df}{dx}V_1\right)V_2
\end{equation}
is obtained.  Similarly, by nesting forward AD inside reverse AD, one can compute
\begin{equation}
  W^T\frac{d}{dx}\left(\frac{df}{dx}V\right).
\end{equation}
Finally, by nesting the stochastic Galerkin scalar type inside the forward AD type, one can obtain the polynomial chaos expansion of the Jacobian from \Eqref{eq:sg_jac}.

Once the relevant portions of an application code have been templated in this manner, a very deep and powerful interface into the computation is provided which can be leveraged for the several analysis tasks discussed above.  By providing a template version of the code, the application developer is essentially creating a generic version of their calculation that can be transformed into a new calculation through template instantiation.  Clearly this template instantiation occurs at compile-time and the corresponding overloaded operators for the chosen scalar type are for the most part incorporated automatically through the compiler's normal template deduction and overload resolution mechanisms.  Thus this approach introduces compile-time polymorphism in each of the mathematical operations invoked in a given calculation through template-based generic programming.

\section{Seed-Compute-Extract Paradigm}
\label{sec:seed-compute-extract}

As was seen in \Secreff{sect:ad}, the forward mode of automatic
differentiation requires seeding the AD objects representing the
independent variables to initialize the AD calculation.
Once seeded, the derivative calculation proceeds by evaluating the
calculation with the scalar type template parameter replaced by the forward AD
type, starting from the seeded independent variables.  When the
calculation arrives at the dependent variables, the AD objects store
the values and derivatives of the dependent variables with respect to
the seeded independent variables (\Eqref{eq:forward_ad}).  Here the
function values and derivative values must be extracted and
communicated to the analysis algorithm requesting them.  This general
paradigm of seeding the objects representing the independent
variables, evaluating the code on these objects by instantiating the
template code on the corresponding types, and extracting the results
is general and applies to all of the analysis types discussed above.

From a generic programming perspective, the only difference between
all of these different evaluations is the seeding and extraction
phase, and these are the only points in the calculation where each
scalar type must be explicitly referenced.  Moreover, the seeding and extraction is for the most part independent of the details of the actual calculation that is being transformed, and only depends on the identification of independent and dependent variables (i.e., which quantities in the program these correspond to and how many of them there are).  We have found it useful to create templated seeding and extraction classes
explicitly specialized for each different type (see, for example,
\cite{CPPTemplatesBook2003} for a discussion on template
specialization).  For a given simulation code, these classes only need to be written once and will suffice for all of the different problems that code simulates, and localizes the scalar type-specific parts of the calculation to one or two classes (for each scalar type).  And since these classes are templated, they can be incorporated into the general templated residual evaluation where the compiler will automatically choose the correct specialization during its template instantiation process.  This structure dramatically simplifies both the addition of new problems or physics and new scalar types for new types of analysis.  In the first case, only the compute parts of the evaluation need to be modified or extended and this will generally be independent of the scalar types.  In the latter case, only new seed/extract classes need to be provided which is independent of the physics or problem dependent parts of the evaluation.  These seed and compute classes provide the linkage between the simulation code evaluation of an analysis quantity (residual, Jacobian, \dots) and the high-level data structures needed for the analysis algorithm (e.g., parallel vector and matrix data structures).  Since these classes are explicitly specialized (and thus are not template code), these high-level data structures themselves do not need to be templated.

From a software engineering perspective, the actual implementation of
the seed-compute-extract paradigm can be rather complex.  In
\Secreff{sec:complexity} we will discuss these issues in greater
detail with respect to the Trilinos tools.

\section{Handling Complexity}
\label{sec:complexity}
The template-based generic programming concepts described above allow
for a rich analysis environment.  However the software engineering
complexity related to implementing these concepts grows quickly.  A
Trilinos package called Phalanx \cite{PhalanxWebSite} was implemented
to simplify the use of template-based generic programming.  While the
main use of this package focuses on partial differential equation
(PDE) assembly, it was written to support any type of evaluation and
is not limited to cell-based discretizations.

In this section we will address three central complexity issues that
arise when evaluating quantites. The first issue deals with the
complexity introduced when supporting a myriad of analysis algorithms.
Each of the algorithms described in \Secreff{sect:analysis} requires
the evaluation of specific object types such as residuals, Jacobians,
and parameter sensitivities.  Providing a flexible and extensible
framework for object evaluation is essential.  The second issue is
complexity associated with the algorithm used to construct
multiphysics models.  In this case, the ability to robustly switch
between different equation sets and constitutive models is critical.
The final issue is complexity associated with data management.
Control of the allocation and placement of data on hardware is
critical for efficient assembly routines.  A flexible system must be
implemented to allow ``power users'' to control data allocation when
possible for specific architectures/hardware.

\subsection{Evaluation Types}
\label{sec:phx::evaluation_types}
The seed-compute-extract paradigm of \Secref{sec:seed-compute-extract}
is a powerful tool for separation of algorithmic analysis requirements
from the implementation of the physics models.  However templating
directly on the scalar type presents a number of difficulties.
Instead, the code base is templated on an {\it{evaluation type}}.  An
evaluation type is a simple \cpp \texttt{struct} that corresponds
directly to an object that is required by the analysis algorithms in
\Secref{sect:analysis}.  For example, residual, Jacobian, parameter
sensitivities, stochastic residual, stochastic Jacobian, and Hessian
are all evaluation types.

By templating on the evaluation type, any method associated with the
evaluation process can be specialized for the particular evaluation
type using template specialization.  This is invaluable when writing
the seed and extract methods.  In a previous implementation, the code
was templated on the scalar type and any specializations based on the
evaluation type were done at runtime using dynamic polymorphism with
abstract base classes.  This added to the complexity of the code by
requiring that a number of objects with pure virtual interfaces be
built and passed through the code whenever a specialization was
needed.  By moving to static polymorphism, the code is much cleaner to
read and debug.  Additionally the multiplicity is handled at compile
time, reducing the overhead of virtual function lookups if the
specialization is nested in a low level loop in the assembly kernel.

By templating on an evaluation type, a scalar type can then be used in
multiple evaluation types.  For example, a Jacobian evaluation and a
parameter sensitivity evaluation can both use a Sacado forward AD
(FAD) object, {\texttt{Sacado::Fad::DFad<double>}}. The difference is
that the seed method and extract method will operate on different
objects.  In the case of a Jacobian evaluation, the FAD objects for
the degrees of freedom will be seeded while for a parameter
sensitivity evaluation, the FAD object for the parameter of interest
will be seeded. In the extract phase, a Jacobian sensitivity
evaluation will extract FAD derivative values into a Jacobian matrix,
while the parameter sensitivity evaluation will extract FAD derivative
values into a vector (or multi-vector for mutiple parameters). If the
code were specialized based on the scalar type, then we could not
differentiate the above calculation without falling back on runtime
polymorphism or partial specialization.

The definition of evaluation types and all supported scalar types is
handled through a single struct in Phalanx.  It is called a ``Traits''
object but instead of implementing a true traits mechanism (see, for
example, \cite{Alexandrescu2001}) it is in effect a type glob to
simplify implementation requirements (future work may include
transitioning this into a true traits mechanism).  This class defines
each evaluation type and binds a default scalar type to the evaluation
type by declaring a typedef for {\texttt{ScalarT}}.  The following
example defines the scalar types and evaluation types for implementing
a residual, a Jacobian, a Jacobian-vector product and a Hessian.

\begin{verbatim}
  struct UserTraits : public PHX::TraitsBase {

    // Scalar Types
    typedef double RealType;
    typedef Sacado::Fad::DFad<RealType> FadType;
    typedef Sacado::Fad::DFad< Sacado::Fad::DFad<RealType> > FadFadType;

    // Evaluation Types with default scalar type
    struct Residual { typedef RealType ScalarT; };
    struct Jacobian { typedef FadType ScalarT;  };
    struct Tangent { typedef FadType ScalarT;  };
    struct Hessian { typedef FadFadType ScalarT;  };
    .
    .
    .
  }
\end{verbatim}

An evaluation is usually associated with one scalar type, but can use
multiple scalar types.  For example a Jacobian calculation normally
uses a Sacado FAD object to compute the derivatives.  However if a
user knows that a certain piece of physics is insensitive to the
degree of freedom, then they can attempt to speed up the code by
replacing the FAD data type with a data type that will not cary out
the derivative computations.

The construction of objects required for an evaluation type is
simplified by using static polymorphism.  We automate the building of
objects by implementing compile time loops over evaluation types using
template metaprogramming techniques \cite{BoostMPLBook2004}
implemented in the Sacado and Boost \cite{BoostLib} libraries.  The user
then only needs to list the evaluation types they would like to build
in the Phalanx Traits struct.

\begin{verbatim}
  struct UserTraits : public PHX::TraitsBase {
    .
    .
    .
    typedef Sacado::mpl::vector<Residual, Jacobian, Tangent, Hessian> EvalTypes;
    .
    .
    .
  }
\end{verbatim}

\subsection{Multiphysics Complexity}
\label{sec:phx:dag}
Multiphysics simulation software is plagued by complexity stemming
from the multitude of models that must be supported in simulating
complex physics systems.  Such software environments must support the
ability to flexibly adapt the physics models as a simulation
transitions to different regimes/problems. This might require the
modification/addition/deletion of equations, constitutive laws and
material models.  Supporting a multiplicity in physics models often
results in complex algorithms and rigid software.  The Phalanx library
implements a new design approach that automates the assembly of an
evaluation type (e.g. residuals and Jacobians) and eliminates the
inherent complexity in physics models.  A thorough description and
analysis of the design is beyond the scope of this paper but can be
found in \cite{DAG-TOMS2011}.  This section will briefly summarize the
design principles and focus on how the template-based generic
programming concepts are applied to the framework.

\subsubsection{The Concept: Data Centric Assembly}
The basic idea is to shift the evaluation of an analysis quantity from
an algorithmic focus (i.e. the properly-ordered sequence of steps to
assemble the quantity of interest) to the concept of data and their
low-level dependencies.  In essence, we are decomposing a complex
problem into a number of simpler problems with managed dependencies.
By exposing data dependencies, an algorithm can be constructed
automatically from the simpler pieces.

This concept is best explained by example.  Here we choose a simple
Continuous Stirred Tank Reactor (CSTR) model of a first-order,
exothermic, irreversible chemical reaction $A\rightarrow B$ in a tank
with ideal mixing and an Arrhenius temperature dependence of the
reaction rate~\cite{uppal:1974}.  Note that this is a trivial example
for demonstration purposes.  In general, balance of either mass or
energy requires
\begin{equation}\label{eq:balance}
  [\mbox{accumulation}] = [\mbox{in}] - [\mbox{out}] + [\mbox{generation}] - [\mbox{consumption}]
\end{equation}
for both the quantity of chemical species $A$ and $B$, and the temperature of the fluid.  If we consider an inlet flow of pure $A$ with concentration $c_{A_f}$ and temperature $T_f$ at a rate $\lambda F$ mixed with a recycle flow with rate of $(1-\lambda)F$ in a reactor of volume $V$ surrounded by a cooling jacket of temperature $T_c$ this becomes
\begin{equation} \label{eq:cstr_full}
  \begin{split}
    V\frac{dc_A}{dt}       &= [F(\lambda c_{A_f} + (1-\lambda)c_A)] - [F c_A] + [0] - [V r_{A\rightarrow B}], \\
    V\rho C_p\frac{dT}{dt} &= [\rho C_pF(\lambda T_f + (1-\lambda)T)]- [\rho C_pFT] + [V(-\Delta H) r_{A\rightarrow B}] - [hA(T-T_c)],
  \end{split}
\end{equation}
where $c_A$ is the concentration of $A$, $r_{A\rightarrow B}$ is the reaction rate, $T$ is the temperature in the tank, $\rho$ is fluid density, $C_p$ is the specific heat, $\Delta H$ is the heat of reaction, $h$ is the heat transfer coefficient to the cooling jacket, and $A$ is the heat transfer area.  For a first-order reaction,
\begin{equation}
\label{eq:reaction_rate}
r_{A\rightarrow B} = k c_A
\end{equation}
 where $k$ is the reaction rate constant.  Typically $k$ is adjusted for the temperature of the medium through the Arrhenius law
\begin{equation}
  k = k_0\exp\left(-\frac{E}{RT}\right),
\end{equation}
where $k_0$ is the pre-exponential factor, $E$ is the activation energy of the reaction, and $R$ is the universal gas constant.

The CSTR model, consisting of two nonlinear ordinary differential equations, can now be defined in the analysis nomenclature of \Secref{sect:analysis}.  The solution vector, $x$, is defined as
\begin{equation}
x = \begin{bmatrix} c_A \\ T \end{bmatrix},
\end{equation}
and the residual vector, $f$, is defined as
\begin{equation}
f = \begin{bmatrix} f_A \\ f_T \end{bmatrix},
\end{equation}
where 
\begin{equation}
\begin{split}
f_A &= V\frac{dc_A}{dt} -F(\lambda c_{A_f} + (1-\lambda)c_A) + F c_A - 0 + V r_{A\rightarrow B}, \\
f_T &= V\rho C_p\frac{dT}{dt} -\rho C_pF(\lambda T_f + (1-\lambda)T) + \rho C_pFT - V(-\Delta H) r_{A\rightarrow B} + hA(T-T_c).
\end{split}
\end{equation}

The next step is to decompose~\Eqref{eq:cstr_full} into a system of
simple pieces.  These simple pieces are called ``evaluators.''
In~\cite{DAG-TOMS2011}, the term ``expression'' was used instead of
evaluator, however in the context of template-based generic
programming, expression templates (an entirely different concept) are
used quite often and so to avoid confusion between expressions and
expression templates, the term evaluator will be used in this paper
and is used in the Phalanx code base.

An evaluator is an atomic unit templated on an evaluation type that
contains the following functionality:
\begin{enumerate}
\item Declares both (1) the fields it evaluates, called {\it{evaluated
      fields}} and (2) the fields it directly depends on to perform
  the evaluation, called {\it{dependent fields}}.  A field is a
  quantity of interest that is computed and stored for use in the
  assembly process.  For example, we might want to compute and store
  the reaction rate, $r_{A\rightarrow B}$, in~\Eqref{eq:cstr_full}.  A
  discussion of fields can be found in section
  \ref{sec:memory_management}.
\item Binds the fields required for the calculation.  When fields are
  available for read/write, a method can be called on an evaluator to
  allow the evaluator to resolve memory for the fields that it writes
  and reads.
\item Calculates the values for the evaluated fields.  When all
  dependent fields that an evaluator requires have been evaluated, an
  evaluator can compute the values of the evaluated fields using the
  values of the dependent fields.
\end{enumerate}

By directly providing the functionality enumerated above, we can automatically
construct algorithms (the properly-ordered sequence of steps to calculate all
required quantities/expressions).  This can be done by providing a few more
abstractions:
\begin{enumerate}
\item A \emph{registry} where all evaluators that may be required for
  the calculation are placed.  This registry allows registration of
  evaluators and provides a way to obtain fully constructed fields via
  a \emph{tag}.  Tags will be discussed in section
  \ref{sec:memory_management}. In the Phalanx library, the registry is
  called a Field Manager and is implemented in the
  {\texttt{PHX::FieldManager}} object.
\item A \emph{memory allocator} used to allocate space to store field
  data.  This will also be discussed in section
  \ref{sec:memory_management}.
\item A mechanism to build the directed acyclic graph (DAG)
  representing the dependency among evaluators.  Once the user defines
  one or more fields that are desired, the required evaluators can be
  resolved from the registry and form the ``root'' nodes in a graph.
  Each node may then be queried to determine its dependencies, which
  are placed on the out-edges of the node.  This process is repeated
  recursively until the ``bottom'' of the graph is reached where there
  are no further out-edges.
\item A \emph{scheduler} that traverses the graph to execute it.  The
  execution graph may be obtained by inverting the dependency graph
  constructed in step 3.  A node is scheduled for execution once all of its
  data-dependents (execution-parents) have completed.  This allows significant
  flexibility in the ordering in which nodes are executed, and facilitates
  task-based parallelism (see \cite{DAG-TOMS2011} for more details).
\end{enumerate}

\begin{figure}
  \begin{center}
    \begin{tikzpicture}[node distance=17mm]

      \node[expr] (RA) {$f_A$};
		  \node[expr] (A_out) [below of=RA] {$f_{A_{out}}$};
		  \node[expr] (A_in) [left of=A_out] {$f_{A_{in}}$};
		  \node[expr] (A_accum) [left of=A_in] {$f_{A_{acc}}$};
      \node[expr] (A_gen) [right of=A_out] {$f_{A_{gen}}$};
      \node[expr] (A_con) [right of=A_gen] {$f_{A_{con}}$};
      \node[expr] (r_AB) [below of=A_con] {$r_{A\rightarrow B}$};
      \node[expr] (k) [below right of=r_AB] {$k$};
      \node[expr] (c_A) [below left of=r_AB] {$c_A$};
      \node[expr] (c_A_dot) [below of=A_in] {$\frac{dc_A}{dt}$};

      \node[expr] (T_gen) [right of=A_con] {$f_{T_{gen}}$};
      \node[expr] (T_con) [right of=T_gen] {$f_{T_{con}}$};
			\node[expr] (T_out) [right of=T_con] {$f_{T_{out}}$};
		  \node[expr] (T_in) [right of=T_out] {$f_{T_{in}}$};
		  \node[expr] (T_accum) [right of=T_in] {$f_{T_{acc}}$};
      	\node[expr] (RT) [above of=T_out] {$f_T$};
      \node[expr] (T) [right of=k] {$T$};
      \node[expr] (T_dot) [below of=T_in] {$\frac{dT}{dt}$};



      \begin{pgfonlayer}{background}
      \path[dep] (RA) edge [bend right=10] (A_accum);
      \path[dep] (RA) edge (A_in);
      \path[dep] (RA) edge (A_out);
      \path[dep] (RA) edge (A_gen);
      \path[dep] (RA) edge [bend left=10] (A_con);
      \path[dep] (A_accum) edge (c_A_dot);
      \path[dep] (A_in) edge (c_A);
      \path[dep] (A_out) edge (c_A);
      \path[dep] (A_con) edge (r_AB);
      \path[dep] (r_AB) edge (k);
      \path[dep] (r_AB) edge (c_A);

      \path[dep] (RT) edge [bend left=10] (T_accum);
      \path[dep] (RT) edge (T_in);
      \path[dep] (RT) edge (T_out);
      \path[dep] (RT) edge [bend right=10] (T_gen);
      \path[dep] (RT) edge (T_con);
      \path[dep] (T_accum) edge (T_dot);
      \path[dep] (T_in) edge (T);
      \path[dep] (T_out) edge (T);
      \path[dep] (T_con) edge (T);
      \path[dep] (T_gen) edge (r_AB);
      \path[dep] (k) edge (T);
      \end{pgfonlayer}

    \end{tikzpicture}
  \end{center}
  \caption{Example Expression dependency graph for the CSTR system~\eqref{eq:cstr_full}.}
  \label{fig:dag}
\end{figure}
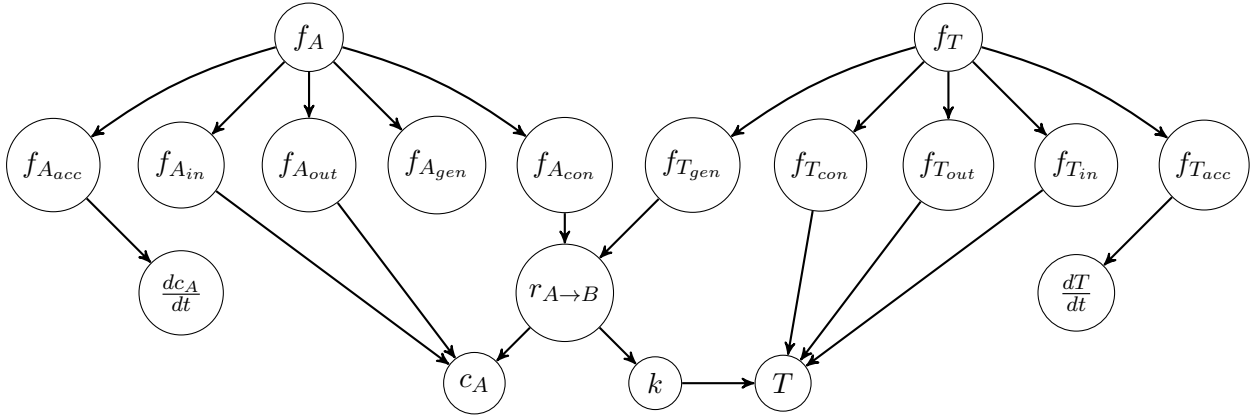

An example decomposition of~\Eqref{eq:cstr_full} is shown
in~\Figref{fig:dag}.
This choice in decompositon follows naturally from separation of terms
in the balance equation~\eqref{eq:balance}.  The residuals are defined as
\begin{equation}
f = \begin{bmatrix} f_A \\ f_T \end{bmatrix} = 
\begin{bmatrix}
f_{A_{acc}} - f_{A_{in}} + f_{A_{out}} - f_{A_{gen}} + f_{A_{cons}} \\
f_{T_{acc}} - f_{T_{in}} + f_{T_{out}} - f_{T_{gen}} + f_{T_{cons}}
\end{bmatrix},
\end{equation}
where the evaluators for fields are defined as
\begin{equation}
\begin{split}
&f_{A_{acc}} = V\frac{dc_A}{dt}, \\
&f_{A_{in}} = F(\lambda c_{A_f} + (1-\lambda)c_A), \\
&f_{A_{out}} =  F c_A, \\
&f_{A_{gen}} =  0, \\
&f_{A_{cons}} =  V r_{A\rightarrow B}, \\
&f_{T_{acc}} = V\rho C_p\frac{dT}{dt}, \\
&f_{T_{in}} = \rho C_pF(\lambda T_f + (1-\lambda)T), \\
&f_{T_{out}} = \rho C_pFT, \\
&f_{T_{gen}} = V(-\Delta H) r_{A\rightarrow B}, \\
&f_{T_{cons}} = hA(T-T_c). \\
\end{split}
\end{equation}
Note that while the accumulation terms are discretized in time using BDF
methods as described in \Secref{sec:transient_analysis}, we recommend this calculation be performed in the high-level time integration package, not within the simulation code itself.   Thus the time derivative terms ($dc_A/dt$ and $dT/dt$) are treated as independent variables in the evaluation of the CSTR residual equations.  
Each node in this graph represents an evaluator and lists
the field(s) that it evaluates.  Out edges from a node point to other
evaluators that evaluate fields that the origin node is directly
dependent on.  Note that the dependency graph nodes are evaluators and
not fields.  This is intentional so that evaluators can evaluate
{\it{multiple}} fields.  This is done to support the incorporation of
third party libraries that, for efficiency, compute multiple fields
(\eg the source terms for the mass and heat balances) in a single
function call.

Note that the decomposition in \Figref{fig:dag} is not unique but is
chosen by the application developers.  All equations could be
implemented in a single model evaluator, or can be split across
multiple evaluators as shown here.  The advantages and disadvantages
of the graph decomposition process with respect to template based
generic programming are discussed in the next section.  A separate
discussion of general advantages irrespective of template based
generic programming can be found in \cite{DAG-TOMS2011}.

\subsubsection{Benefits of Graph-based Assembly}
The use of a graph-based approach has a number of benefits. First, the
evaluation of terms will be consistent.  If a field is dependent on
another field, it will not be calculated until all of its dependent
fields are first calculated.  For a simple problem such as described
by~\Eqref{eq:cstr_full} it is not difficult.  The logical order is
clear and can be implemented by visual inspection.  However as a code
allows for multiple equations and closure models to be added,
subtracted, or swapped, the complexity makes visual inspection and
ordering nearly impossible for all combinations and thus results in
fragile code.  This graph-based approach eliminates the logical
complexity.

The use of fields and graph-based evaluation also provides a
mechanism for code reuse.  As different models are inserted into the
graph and the dependency tree changes, the order is automatically
updated and the surrounding nodes can be reused. Changing dependencies
on one node does not require a code change in any of the surrounding
nodes.  For example, a simpler model of the reaction constant $k$ can
be used that doesn't depend on the temperature $T$.  This can easily
be accomplished by using a different evaluator for $k$ without
changing any other parts of the code\footnote{Note however we assume the dependencies of any given evaluator are static throughout the computational simulation.  Thus branching conditions are allowed within the implementation of any given evaluator as long as they don't change the structure of the graph.  
This latter case could be supported by allowing dynamic reconfiguring of the graph, however our current tools do not support this.}. 
Thus the use of evaluators
isolates model code.

The dependency graph can generate very efficient code.  Due to the
complexity of ever changing equations and closure models, some
applications ignore the complexity and resort to recomputing the same
quantities over and over if needed in separate equation sets and/or
operators, resulting in highly inefficient code.  By using the
dependency graph, a field is guaranteed to be computed once and used
by all other evaluators that depend on it.  For example, in
\Figref{fig:dag}, $r_{A\rightarrow B}$ is computed once and is used by
both the $A$-consumption and $T$-generation evaluators.  Additionally
once $r_{A\rightarrow B}$ is used by all dependent evaluators, the
scheduler could even recycle the memory for $r_{A\rightarrow B}$ to be
used for other fields not yet processed in the DAG.

In terms of template-based generic programming, each evaluator is a
separate \cpp object templated on the evaluation type.  As such, it
not only becomes a point of variability to switch models, it can also
be a point of variability in evaluation types by using template
specialization on a single model.  The user can implement different
algorithms based on the evaluation type for each model.  For example,
in some cases having an exact Jacobian can hinder convergence rates.
The {\texttt{Jacobian}} evalution type could be specialized to drop
certain dependencies from the AD scalar types.

The ability to apply template specialization to evaluators for a
particular evaluation type is critical in the seed-compute-extract
paradigm in \Secref{sec:seed-compute-extract}.  For each of the
evaluation types, only the seed and extract evaluators need be
specialized.  All other evaluators in the graph are used in the
compute phase and have a {\it{single implementation templated on the
    scalar type}}.

\subsection{Data Structures and Memory Management}
\label{sec:memory_management}
A critical aspect to efficient assembly algorithms is the management
of data.  This section is devoted to discussing a number of issues
that impacted the design of Phalanx.

The allocation of memory for storing field data in the Phalanx library
is completely controlled by the user.  A templated parameter in the
Traits class is used to specify a memory allocator object that can be
written by the user.  Once the dependency graph is constructed, the
registry contains a list of all fields that the registry requires
memory for.  The registry can then call the allocator to dynamically
allocate storage for the field data.  An important aspect is that the
allocator could allocate a single contiguous array for all field data
members (including mixed data types) in an attempt to fit all field
data for a complete graph traversal into cache.  Phalanx contains two
allocator implementations.  A simple allocator that uses the \cpp
{\texttt{new}} command and an advanced allocator that allocates one
single contiguous block of memory for all fields of all data types.
The advanced allocator is templated on an alignment type so that the
arrays can be correctly aligned for a base scalar type.  When a block
of the contiguous array is assigned to a field, a special reinterpret cast is called on the
block to correctly call the scalar type constructors.  This
functionality along with the reference counting guarantees that the
constructors and destructors of the field scalar type are called
correctly.  However for flexibility, users are free to specify their own allocator.  This is useful, for example, when users have their own mesh database for storing field data.

%
%
%

The field interface is a simple handle layer.  A field consists of a
tag used for identification in the registry and a reference-counted
smart pointer (RCP) \cite{Bartlett2010,Alexandrescu2001} for accessing
the field data.  Two implementations of fields are available in
Phalanx, a simple {\texttt{PHX::Field}} object and the more advanced
multidimensional array {\texttt{PHX::MDField}} object.  While the
memory management relies on RCPs, in the future the array will
optionally support out-of-core memory for running on GPGPUs using the
multidimensional array under development in the Kokkos package
\cite{KokkosWebSite}.

The tag on a field is used to provide a unique signature for
identifying fields in the registry.  It is constructed from a name (a
{\texttt{std::string}}), a scalar type, and a data layout.  The scalar
type information is returned in a {\texttt{std::type\_info}}
object to support runtime comparisons.  The data layout object
specifies the data signature for a multidimensional array.  It
contains the rank of the array, the names of the ordinals in the
array, the sizes of the ordinals in the array, and an optional name.
The unique signature is formed from a combination of these three
objects so fields can be differentiated by name, scalar type, or data
layout.

\subsection{An Example Evaluator}
Now that the data structures and algorithms have been discussed, an
example evaluator for the node representing the reaction rate
$r_{A\rightarrow B}$, will be shown.  The header file, excluding the include
guards and (optional) explicit template instantiation declarations, is
shown in \Figref{fig:evaluator_header}.
\begin{figure}
\begin{verbatim}
#include "Phalanx_ConfigDefs.hpp"
#include "Phalanx_Evaluator_WithBaseImpl.hpp"
#include "Phalanx_Evaluator_Derived.hpp"
#include "Phalanx_MDField.hpp"

template<typename EvalT, typename Traits>
class FirstOrderReaction : public PHX::EvaluatorWithBaseImpl<Traits>,
		public PHX::EvaluatorDerived<EvalT, Traits>  {

public:
  FirstOrderReaction(const Teuchos::ParameterList& p);
  void postRegistrationSetup(typename Traits::SetupData d,
			     PHX::FieldManager<Traits>& vm);
  void evaluateFields(typename Traits::EvalData d);

private:

  typedef typename EvalT::ScalarT ScalarT;

  PHX::MDField<ScalarT> k;      // reaction rate constant
  PHX::MDField<ScalarT> c;      // concentration
  PHX::MDField<ScalarT> r;      // total reaction rate
}
\end{verbatim}
\caption{Header file source for implementing the reaction rate $r_{A\rightarrow B}$.  Note that the include guards and (optional) explicit template instantiation machinery are not shown for clarity.}
\label{fig:evaluator_header}
\end{figure}
Three fields are declared.  One for the reaction rate values,
$r_{A\rightarrow B}$, and two for the dependent fields, $k$ and $c_A$,
required to perform the calculation in \Eqref{eq:reaction_rate}.  A set of macros can be used to hide the
class declaration boilerplate.  This substantially cuts down on the
boilerplate and additionally hides most of the templating from users.
The same header from \Figref{fig:evaluator_header} is shown in
\Figref{fig:evaluator_header_simplified} using the macros.

\begin{figure}
\begin{verbatim}
#include "Phalanx_ConfigDefs.hpp"
#include "Phalanx_Evaluator_WithBaseImpl.hpp"
#include "Phalanx_Evaluator_Derived.hpp"
#include "Phalanx_MDField.hpp"

PHX_EVALUATOR_CLASS(FirstOrderReaction)

  PHX::MDField<ScalarT> k;      // reaction rate constant
  PHX::MDField<ScalarT> c;      // concentration
  PHX::MDField<ScalarT> r;      // total reaction rate

PHX_EVALUATOR_CLASS_END
\end{verbatim}
\caption{Header file source for implementing the reaction rate $r_{A\rightarrow B}$ using macro definition for class declaration boilerplate.}
\label{fig:evaluator_header_simplified}
\end{figure}

The implementation of the three methods for this evaluator class follows.  The constructor, the post registration setup and the evaluate methods are shown in \Figref{fig:evaluator_definition_ctor} using the macro definitions.
\begin{figure}
\begin{verbatim}
PHX_EVALUATOR_CTOR(FirstOrderReaction,params)
{
  Teuchos::RCP<PHX::DataLayout> scalar =
    params.get<Teuchos::RCP<PHX::DataLayout> >("Scalar Layout");

  k = PHX::MDField<ScalarT>("k", scalar);
  c = PHX::MDField<ScalarT>("c", scalar);
  r = PHX::MDField<ScalarT>("r", scalar);

  this->addDependentField(k);
  this->addDependentField(c);
  this->addEvaluatedField(r);
}

PHX_POST_REGISTRATION_SETUP(FirstOrderReaction,user_data,field_manager)
{
  this->utils.setFieldData(k,field_manager);
  this->utils.setFieldData(c,field_manager);
  this->utils.setFieldData(r,field_manager);
}

PHX_EVALUATE_FIELDS(FirstOrderReaction,user_data)
{
   r(0) = k(0) * c(0);
}
\end{verbatim}
\caption{Evaluator implementation.}
\label{fig:evaluator_definition_ctor}
\end{figure}
The constructor is used to create the tag for each field.  This
consists of a string name and the associated data layout.  The data
layout is passed in by the user through a parameter list.  The data
layout contains a description of the multidimensional array as
described in \Secref{sec:memory_management}.  In the constructor, the
evaluator is told how to advertise its capabilities.  It declares what
fields it evaluates and what fields it requires to complete the
evaluation.  The post registration setup function is used for binding
the memory allocated for each field.  The {\texttt{field\_manager}}
object is the field manager and contains the registry with pointers
for the memory for each field.

The evaluate routine is the actual implementation of the algorithm.
While this is a very simple expression, note that there is no limit on
how much work a single operator could do.  For example, a single
evaluator could implement multiple complete equations instead of this
single term.

The {\texttt{PHX::MDField}} object is a multidimensional array that
wraps the memory associated with storing the values of a field.  In
the case of our simple demonstration problem the fields are single
scalar values (i.e. a rank one array of size one).  This notation is
slightly clunky for single scalars since one would prefer writing
{\texttt{r = k * c}} instead of {\texttt{r(0) = k(0) * c(0)}}.
However, most use cases require MDFields that are larger arrays than
single scalars (such as for finite element calculations).  In such
cases, multidimensional array support is crucial for performance. Note
that one could recover the simple syntax by implementing expression
templates for the MDField objects.



\section{Numerical Examples} \label{sect:example}
We now provide simple numerical examples that demonstrate the power of this template-based generic programming approach.  While these ideas can be applied to very complex multiphysics systems, for pedagogical purposes we consider the simple CSTR equations~\eqref{eq:cstr_full}.  Using the ideas discussed above, Phalanx evaluators can be written for each of the terms appearing in \Eqref{eq:cstr_full}, along the lines of those shown in Figs.~\ref{fig:evaluator_header}--\ref{fig:evaluator_definition_ctor}.  These evaluators, along with seed and extract evaluators for each evaluation type and interfaces connecting these calculations to high-level simulation and analysis algorithms in Trilinos, provide the software implementation necessary for the numerical examples shown below.  We present these examples in terms of the non-dimensional version~\cite{uppal:1974} of \Eqref{eq:cstr_full}:
\begin{equation}\label{eq:cstr}
  \begin{split}
    \frac{dx}{dt} &= -x + D(1-x)\exp\left(\frac{y}{1+y/\gamma}\right), \\
    \frac{dy}{dt} &= -y + BD(1-x)\exp\left(\frac{y}{1+y/\gamma}\right)-\beta(y-y_c).
  \end{split}
\end{equation}
These equations exhibit a variety of interesting dynamical phenomena including multiple stable steady-states, stable and unstable oscillations, and bifurcations depending on the values of the parameters $D$, $B$, $\gamma$, $\beta$ and $y_c$.  As in~\cite{uppal:1974}, we consider the case $\gamma\rightarrow\infty$ and $y_c=0$.

\begin{figure}[H]
  \centering
  \subfloat[Continuation.]{\label{fig:cont}\includegraphics[scale=0.35]{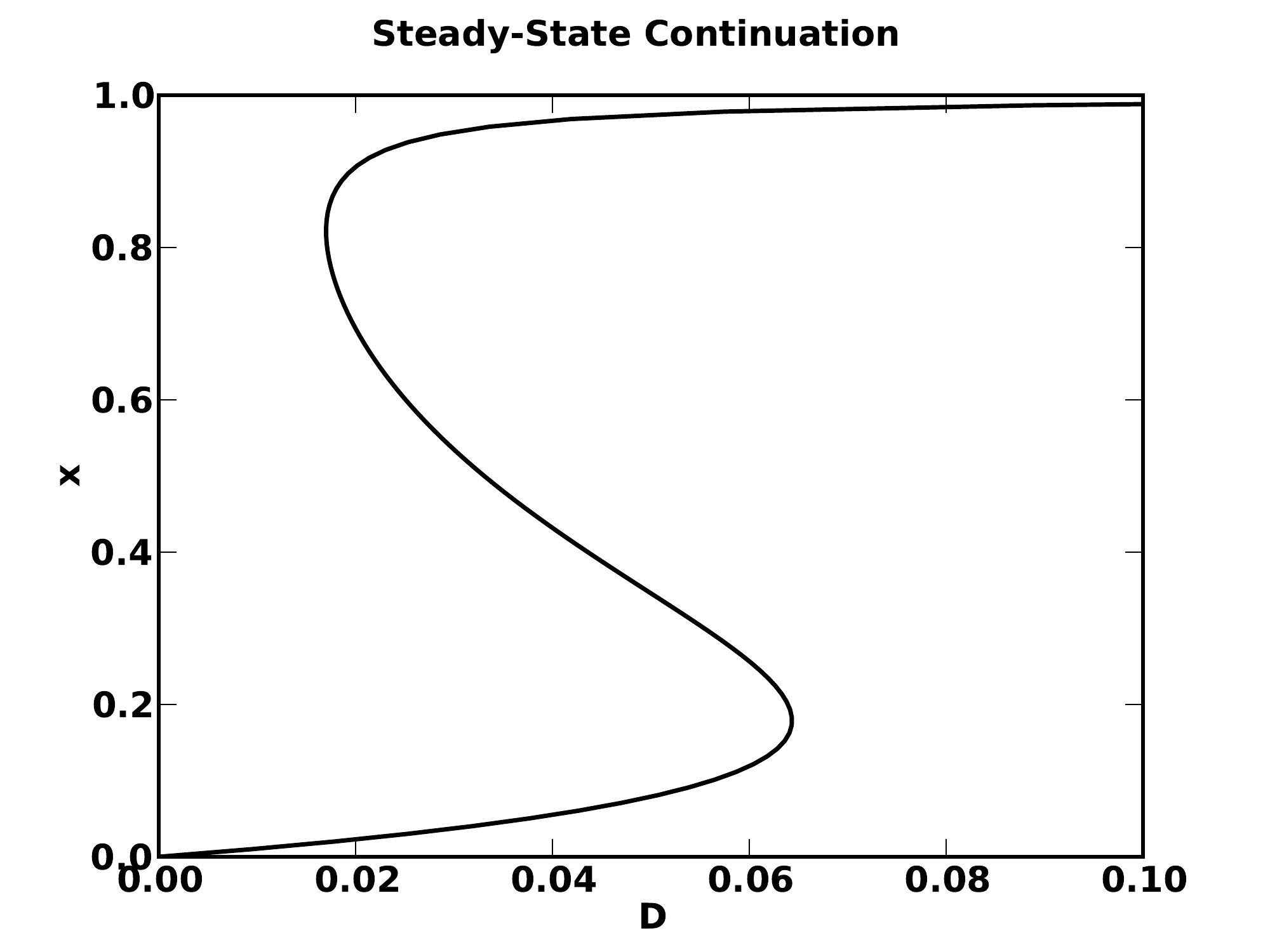}}
  \subfloat[Turning-point continuation.]{\label{fig:tp}\includegraphics[scale=0.35]{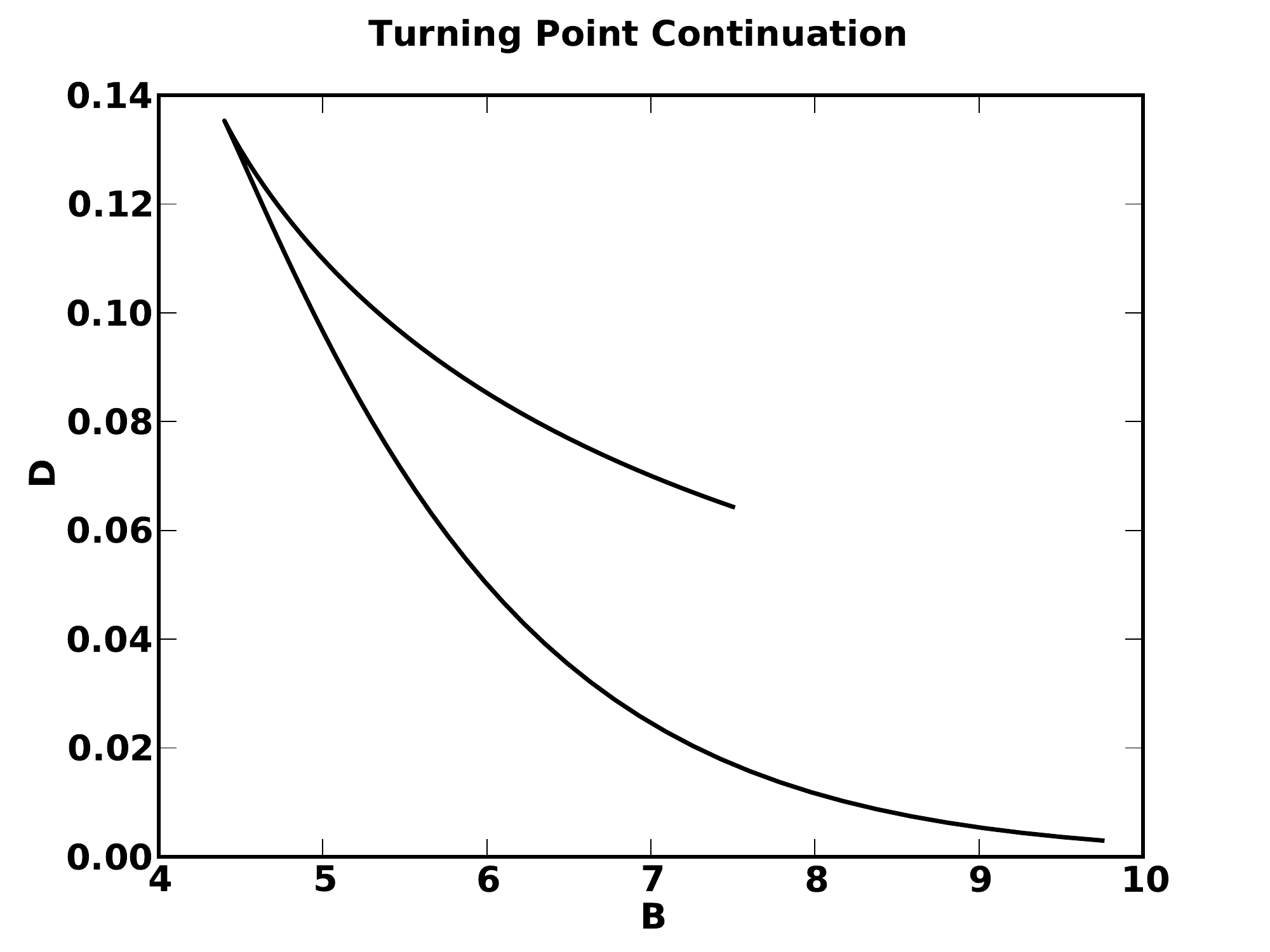}} \\
  \subfloat[Transient]{\label{fig:trans}\includegraphics[scale=0.35]{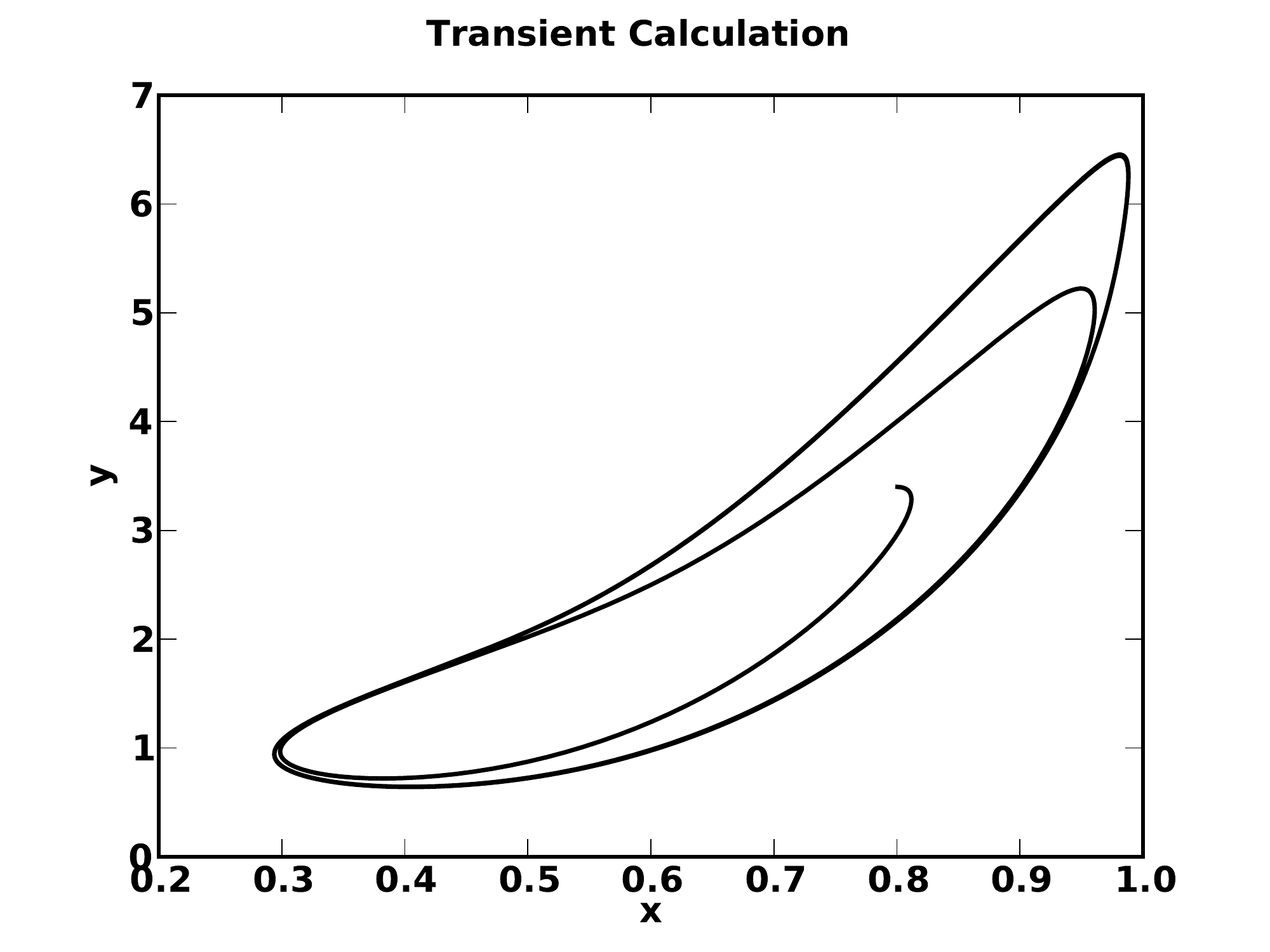}} \\
  \subfloat[$x$ Density]{\label{fig:density_x}\includegraphics[scale=0.35,page=1]{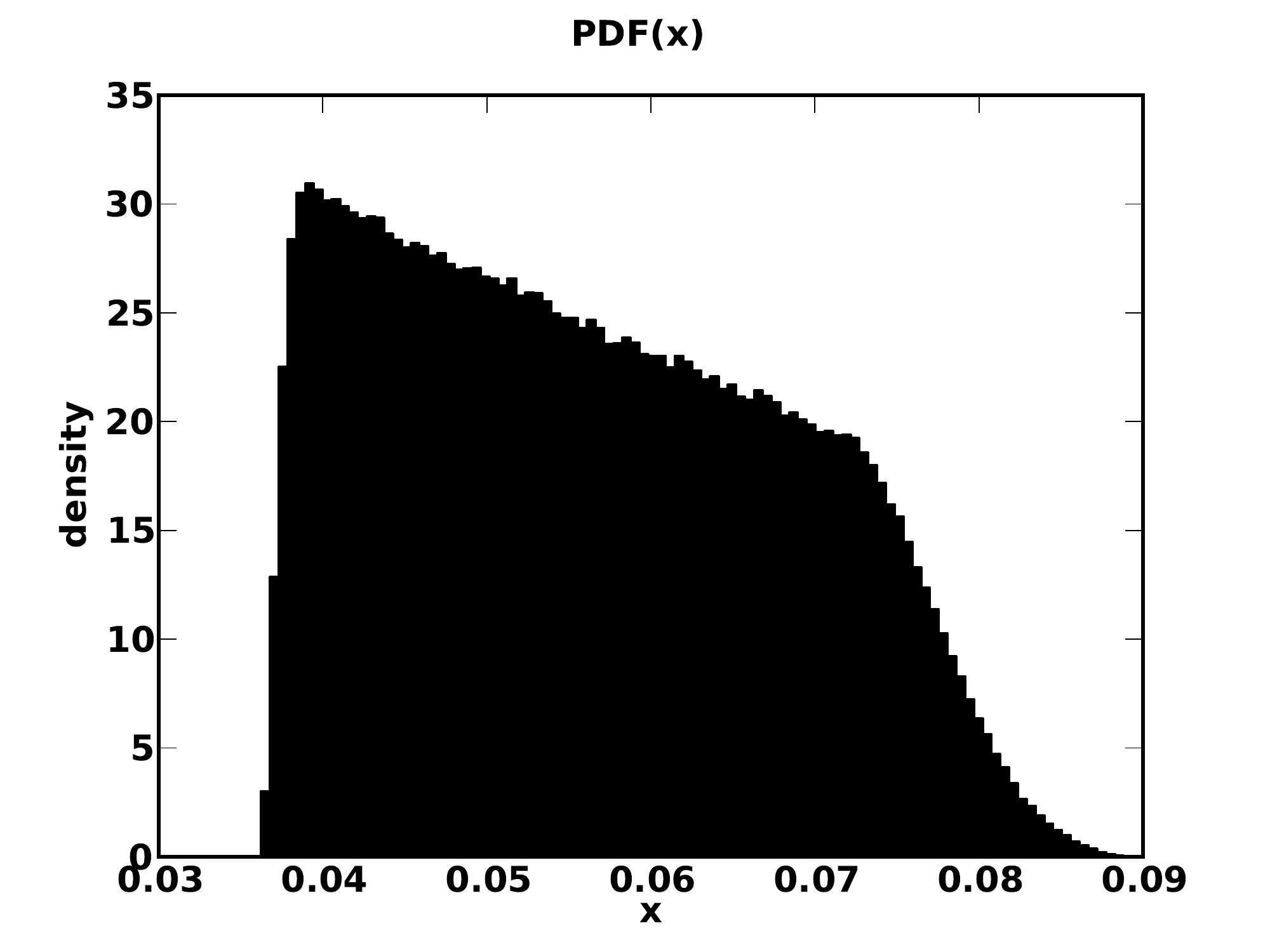}}
  \subfloat[$y$ Density]{\label{fig:density_y}\includegraphics[scale=0.35,page=2]{density}}
\caption{Numerical examples of several analysis approaches applied to the CSTR equations}
\label{fig:example}
\end{figure}

We first consider the locus of steady-state solutions as a function of $D$ for fixed $B$ and $\beta$.  Such a curve is shown in \Figref{fig:cont}, which was computed using the LOCA~\cite{LocaURL} and NOX~\cite{NoxURL} packages in Trilinos {\em via} pseudo-arclength continuation.  This curve illuminates values of $D$ where multiple steady-states exist and the existence of turning-point bifurcations.  The parameter values where these bifurcations occur can be solved for using the bifurcation equations~(\ref{eq:bif},~\ref{eq:sigma_def}) and tracked as a function of a second parameter (also through pseudo-arclength continuation).  Such a calculation using LOCA is displayed in \Figref{fig:tp}.  For both of these calculations, the template-based generic programming ideas discussed above are leveraged for analytic evaluation of the Jacobian and parameter derivatives needed by these algorithms as discussed in \Secref{sect:stability}.  Note that the implementation of these equations is capable of providing the analytic second derivatives needed by the turning point algorithm~\eqref{eq:sigma_derivs}, and LOCA is capable of using them, however the interface used to connect this problem to LOCA currently doesn't support them and thus they were estimated by first-order finite differencing of the first derivatives.

For other choices of the model parameters, the system exhibits oscillations.  An example of this is shown in \Figref{fig:trans} using the Rythmos time integration package~\cite{RythmosURL} in Trilinos.  In this case, analytic derivatives with respect to the transient terms are provided through the template-based generic programming approach and the Sacado automatic differentiation package.

As a final example, we consider the case when $D$, $B$, and $\beta$ are uncertain as represented by the following independent random variables:
\begin{equation}
  \begin{split}
    D &\sim \mbox{uniform on} \; [0.03,0.05], \\
    B &\sim \mbox{uniform on} \; [7,8], \\
    \beta &\sim \mbox{uniform on} \; [0.05,1.05].
  \end{split}
\end{equation}
We then estimate a fourth order polynomial chaos expansion of the steady-state solution to \Eqref{eq:cstr} using the intrusive stochastic Galerkin approach from \Secref{sec:uq}.  In this case the polynomials $\psi$ are tensor products of one dimensional Legendre polynomials.  The stochastic Galerkin residual equations~\eqref{eq:sg_resid} and corresponding Jacobians~\eqref{eq:sg_jac} are evaluated using the specialized overloaded operators discussed above using the Sacado and Stokhos packages.  These are then assembled into the stochastic Galerkin linear systems needed to solve the equations using Newton's method as provided by NOX~\cite{NoxURL}.  Once the solution is computed, it can be employed for a variety of analysis purposes.  For example, in \Figref{fig:density_x} and~\ref{fig:density_y} we plot an estimation of the probability density function for $x$ and $y$, computed by Monte Carlo sampling of the polynomial chaos response surface $x(\xi)$, $y(\xi)$.

The purpose of these calculations is not to demonstrate any new numerical results or new algorithmic approaches, rather to demonstrate how a single templated code base can be leveraged for a variety of analysis algorithms that require advanced capabilities of the simulation code.  While all of these calculations are very intrusive to the simulation code, the use of template-based generic programming for the most part makes these calculations transparent to the code developer and provides hooks for incorporation of future intrusive analysis algorithms.

\section{Concluding remarks}\label{sect:conclusions}

In this paper we described a template-based generic programming approach for incorporating advanced analysis algorithms into complex simulation codes.  This approach relies on operator overloading techniques in the \cpp language inspired by automatic differentiation to transform a base residual calculation into one that can compute a variety of derivative and non-derivative quantities useful for advanced simulation and analysis.  The approach also leverages the \cpp template facilities to automate the conversion from the base scalar floating point type to the multitude of scalar types needed for operator overloading.  Additionally we described a graph-based assembly approach for creating complex simulation tools, and how this approach builds on template-based generic programming.  While these techniques are general and could be implemented in a variety of software tools, we described several tools within the Trilinos software framework that implement these ideas, namely the Sacado automatic differentiation, Stokhos stochastic Galerkin, and Phalanx graph-based assembly packages.  Together with advanced solver and analysis tools in Trilinos such as NOX, LOCA, and Rythmos, as well as other simulation tools such as geometry specification, mesh database, and discretization tools also available in Trilinos, simulation tools for complex physical, biological, and engineered processes can be developed that support a wide variety of state-of-the-art simulation and analysis algorithms.  These tools and techniques have already been applied to more interesting and challenging problems than the simple example described here, which is the subject of a subsequent paper~\cite{Salinger:2011}.  Together these papers demonstrate the most compelling aspect of the overarching template-based generic programming approach:  while it is always possible to develop a customized software tool for a given kind of simulation or analysis (e.g., transient, stability, or uncertainty analysis), the approach allows a {\em single} software implementation of the underlying equations to be leveraged for all of these analyses.  

We admit that the approach described here is complex and requires a fair degree of sophistication on the part of the programmer in terms of familiarity with templating, polymorphism, and operator overloading.  However the beauty of the template-based generic programming approach is that it for the most part allows the programmer to ignore, and perhaps not even be aware of, the embedded analysis that ultimately might be performed on the simulation or the details of the specific scalar types and their corresponding overloaded operators.

\begin{center}
\bf\large Acknowledgements
\end{center}

This work was funded by the US Department of Energy through the NNSA Advanced Scientific Computing and Office of Science Advanced Scientific Computing Research programs.

\bibliographystyle{abbrvnat}
\bibliography{paper}

\end{document}

%% file: symbol_defs.tex
\renewcommand{\vector}[1]{\ensuremath{{\boldsymbol{#1}}}}

\newcommand{\x}{\vector{x}}

